\documentclass[english,showpacs,preprint,superscriptaddress]{revtex4-1}
\usepackage{amsmath}
\usepackage{amssymb}
\usepackage{graphicx}
\usepackage{esint}

\makeatletter

\providecommand{\tabularnewline}{\\}

\usepackage{amsthm}\usepackage{latexsym}\usepackage{bm}\usepackage{amsfonts}

\usepackage{babel}

\usepackage{hyperref}

\hypersetup{
	breaklinks = true,
    colorlinks = true,
    citecolor = {blue},
	urlcolor = {blue},
	linkcolor = {blue}
}

\usepackage{babel}

\global\long\def\EXP{\times10^}
\usepackage{cleveref}
\global\long\def\nbfha{{h}_{\sigma1,x_i}}
 
 \global\long\def\nbfhc{{h}_{\sigma3,x_i}}
 \global\long\def\diag{\textrm{diag}}

 \global\long\def\rmd{\mathrm{d}}
 \global\long\def\rmc{\mathrm{c}}
 \global\long\def\rmm{\mathrm{m}}
 \global\long\def\diag{\textrm{diag}}
 
 \global\long\def\bfx{\mathbf{x}}
 
 \global\long\def\bfv{\mathbf{v}}
 \global\long\def\bfha{\mathbf{h}_{\sigma1}}
 \global\long\def\bfhb{\mathbf{h}_{\sigma2}}
 \global\long\def\bfhc{h_{\sigma3}}
 \global\long\def\bfA{\mathbf{A}}
 \global\long\def\bfJ{\mathbf{j}}
 \global\long\def\bfB{\mathbf{B}}

 \global\long\def\bfE{\mathbf{E}}

 \global\long\def\bfe{\mathbf{e}}

 \global\long\def\FIG#1{Fig.~\ref{#1}}
 \global\long\def\EQ#1{Eq.~(\ref{#1})}
 \global\long\def\SEC#1{Sec.~\ref{#1}}
 \global\long\def\APP#1{Appendix~\ref{#1}}
 \global\long\def\REF#1{Ref.~\citep{#1}}

 \global\long\def\CURLD{ {\mathrm{curl_{d}}}}
 \global\long\def\DIVD{ {\mathrm{div_{d}}}}
 \global\long\def\CURLDP{ {\mathrm{curl_{d}}^{T}}}

 \global\long\def\calA{\mathcal{A}}

\newcommand{\WZERO}[1]{W_{\sigma_0 J}\left( #1 \right)}
\newcommand{\WONE}[1]{W_{\sigma_1 J}\left( #1 \right)}

\newcommand{\WTWO}[1]{W_{\sigma_2 J}\left( #1 \right)}
\newcommand{\WTHREE}[1]{W_{\sigma_3 J}\left( #1 \right)}

\newcommand{\WOZERO}[1]{W_{O\sigma_0 J}\left( #1 \right)}
\newcommand{\WOONE}[1]{W_{O\sigma_1 J}\left( #1 \right)}

\newcommand{\WOTWO}[1]{W_{O\sigma_2 J}\left( #1 \right)}
\newcommand{\WOTHREE}[1]{W_{O\sigma_3 J}\left( #1 \right)}

\usepackage{babel}

\usepackage{babel}

\usepackage{babel}

\@ifundefined{showcaptionsetup}{}{%
 \PassOptionsToPackage{caption=false}{subfig}}
\usepackage{subfig}
\makeatother

\begin{document}

\title{Explicit Structure-Preserving Geometric Particle-in-Cell Algorithm
in Curvilinear Orthogonal Coordinate Systems and Its Applications
to Whole-Device 6D Kinetic Simulations of Tokamak Physics}

\author{Jianyuan Xiao}
\email{xiaojy@ustc.edu.cn}

\affiliation{School of Nuclear Science and Technology, University of Science and
Technology of China, Hefei, 230026, China}

\author{Hong Qin}
\email{hongqin@princeton.edu}

\affiliation{Plasma Physics Laboratory, Princeton University, Princeton, NJ 08543,
U.S.A.}
\begin{abstract}
Explicit structure-preserving geometric Particle-in-Cell (PIC) algorithm
in curvilinear orthogonal coordinate systems is developed. The work
reported represents a further development of the structure-preserving
geometric PIC algorithm \citep{squire4748,squire2012geometric,xiao2013variational,xiao2015explicit,xiao2015variational,he2015hamiltonian,qin2016canonical,he2016hamiltonian,kraus2017gempic,xiao2017local,xiao2018structure,xiao2019field},
achieving the goal of practical applications in magnetic fusion research.
The algorithm is constructed by discretizing the field theory for
the system of charged particles and electromagnetic field using Whitney
forms, discrete exterior calculus, and explicit non-canonical symplectic
integration. In addition to the truncated infinitely dimensional symplectic
structure, the algorithm preserves exactly many important physical
symmetries and conservation laws, such as local energy conservation,
gauge symmetry and the corresponding local charge conservation. As
a result, the algorithm possesses the long-term accuracy and fidelity
required for first-principles-based simulations of the multiscale
tokamak physics. The algorithm has been implemented in the \emph{SymPIC}
code, which is designed for high-efficiency massively-parallel PIC
simulations in modern clusters. The code has been applied to carry
out whole-device 6D kinetic simulation studies of tokamak physics.
A self-consistent kinetic steady state for fusion plasma in the tokamak
geometry is numerically found with a predominately diagonal and anisotropic
pressure tensor. The state also admits a steady-state sub-sonic ion
flow in the range of 10 km/s, agreeing with experimental observations
\citep{Ince-Cushman2009,Rice2009} and analytical calculations \citep{Guan2013,Guan2013a}.
Kinetic ballooning instability in the self-consistent kinetic steady
state is simulated. It shows that high-n ballooning modes have larger
growth rates than low-n global modes, and in the nonlinear phase the
modes saturate approximately in 5 ion transit times at the 2\% level
by the $E\times B$ flow generated by the instability. These results
are consistent with early \citep{Qin98-thesis,Qin1999} and recent
\citep{dong2019nonlinear} electromagnetic gyrokinetic simulations. 
\end{abstract}
\maketitle

\section{Introduction}

\label{sec:Introduction}Six Dimensional (6D) Particle-In-Cell (PIC)
simulation is a powerful tool for studying complex collective dynamics
in plasmas \citep{Dawson1983,hockney1988computer,birdsall1991plasma}.
However, as a first-principles-based numerical method, 6D PIC simulation
has not been employed to study the dynamical behavior experimentally
observed in magnetic fusion plasmas, mainly due to the multiscale
nature of the physics involved. The technical difficulties are two-fold.
First, the span of space-time scales of magnetic fusion plasmas is
enormous, and numerically resolving these space-time scales demands
computation hardware that did not exist until very recently. Secondly,
even if unlimited computing resource exists, the standard 6D PIC algorithms
do not posses the long-term accuracy and fidelity required for first-principles-based
simulations of magnetic fusion plasmas. For example, the most commonly
used 6D PIC scheme is the Boris-Yee scheme, which solves the electromagnetic
fields using the Yee-FDTD \citep{yee1966numerical} method and advances
particles' position and velocity by the Boris algorithm \citep{Boris70}.
Even though both Yee-FDTD method \citep{stern2007geometric} and the
Boris algorithm \citep{qin2013boris,he2015volume,zhang2015volume,ellison2015comment,He16-172,Tu2016}
themselves do have good conservative properties, their combination
in PIC methods does not \citep{hockney1988computer,birdsall1991plasma,ueda1994study}.
Numerical errors thus accumulate coherently during simulations and
long-term simulation results are not trustworthy.

Recent advances in super-computers \citep{fu2016sunway} have made
6D PIC simulation of magnetic fusion plasmas possible in terms of
hardware. On the other hand, much needed now is modern 6D PIC algorithms
that can harness the rapidly increasing power of computation hardware.
A family of such new 6D PIC algorithms \citep{squire4748,squire2012geometric,xiao2013variational,xiao2015explicit,xiao2015variational,he2015hamiltonian,qin2016canonical,he2016hamiltonian,kraus2017gempic,xiao2017local,xiao2018structure,xiao2019field}
has been developed recently. Relative to the conventional PIC methods,
two unique properties characterize these new PIC algorithms: the space-time
discretization is designed using modern geometric methods, such as
discrete manifold and symplectic integration, and the algorithms preserve
exactly many important physical symmetries and conservation laws in
a discrete sense. For this reason, these algorithms are called structure-preserving
geometric PIC algorithms. The long-term accuracy and fidelity of these
algorithms have been studied and documented by different research
groups \citep{xiao2015explicit,kraus2017gempic,Morrison2017,Holderied2020,xiao2019commet,Li2019,Li2020,hirvijoki2020,Perse2020,Zheng2020,wang2020geometric,Kormann2021}.

Discrete symplectic structure and symplectic integration are two of
the main features of the structure-preserving geometric PIC algorithms.
Up to now, the only explicit symplectic integrator for the non-canonical
symplectic structure of the discrete or continuous Vlasov-Maxwell
system is the splitting method developed in Ref.\,\citep{he2015hamiltonian},
which splits the dynamics symplectically into one-way decoupled Cartesian
components. The requirement of the Cartesian coordinate system creates
a hurdle for applications in magnetic fusion plasmas, for which cylindrical
or toroidal coordinate systems are more convenient. Most simulation
codes, such as GTC \citep{lin1998turbulent,zebin2013gtc}, XGC \citep{ku2006gyrokinetic,chang2009compressed}
and GEM \citep{chen2003deltaf,chen2007electromagnetic,wang2012linear}
adopted curvilinear coordinate systems. In the present study, we extend
the structure-preserving geometric PIC algorithm to arbitrary curvilinear
orthogonal coordinate systems, in particular, to the cylindrical coordinate
system for applications in tokamak physics. We show that when a sufficient
condition (\ref{EqnH123}) is satisfied, which is the case for the
cylindrical mesh, the structure-preserving PIC algorithm can be made
explicitly solvable and high-order. For easy implementation and presentation,
we first present the algorithm as a discrete field theory \citep{lee82,lee87,veselov88,marsden2001discrete,Guo2002,Qin2019}
specified by a Lagrangian discretized in both space and time. It is
then reformulated using a generalized version of the discrete Poisson
bracket developed in Ref.\,\citep{xiao2015explicit} and an associated
explicit symplectic splitting algorithm, which generalizes the original
algorithm designed in Ref.\,\citep{he2015hamiltonian} in the Cartesian
coordinate system. In particular, the technique utilizing the one-way
decoupling of dynamic components is generalized to arbitrary curvilinear
orthogonal coordinate systems satisfying condition (\ref{EqnH123}).

The algorithm has been implemented in the the\textsl{ SymPIC} code
and applied to carry out whole-device 6D kinetic PIC simulations of
plasma physics in a small tokamak with similar machine parameters
as the Alcator C-Mod \citep{hutchinson1994first,greenwald1997h}.
We numerically study two topics: self-consistent kinetic steady state
in the tokamak geometry and kinetic ballooning instability in the
edge. Simulation shows that when plasma reaches a self-consistent
kinetic steady state, the pressure tensor is diagonal, anisotropic
in 3D, but isotropic in the poloidal plane. A sub-sonic ion flow in
the range of $10$km/s exists, which agrees with experimental observations
\citep{Ince-Cushman2009,Rice2009} and theoretical calculations \citep{Guan2013,Guan2013a}.
In the self-consistent kinetic steady state, it is found that large-$n$
kinetic ballooning modes grow faster than low-$n$ global modes, and
the instability saturates nonlinearly at the 2\% level by the $E\times B$
flow generated by the instability. These results qualitatively agree
with previous simulations by electromagnetic gyrokinetic codes \citep{Qin98-thesis,Qin1999,dong2019nonlinear}.

Because the algorithm is based on first-principles of physics and
constructed in general geometries, the whole-device 6D kinetic simulation
capability developed is applicable to other fusion devices and concepts
as well, including stellarators, field reserve configurations and
inertial confinement. It is also worthwhile to mention that the structure-preserving
geometric discretization technique developed for the 6D electromagnetic
PIC simulations can be applied to other systems as well, including
ideal two-fluid systems \citep{xiao2016explicit} and magnetohydrodynamics
\citep{zhou2014variational,zhou2015formation,zhouthesis,Zhou2017APJ,burby2017a}.
Structure-preserving geometric algorithms have also been developed
for the Schrödinger-Maxwell \citep{chen2017canonical} system, the
Klein-Gordon-Maxwell system \citep{Shi2016,Shi2018,ShiThesis,Shi2020},
and the Dirac-Maxwell system \citep{Chen2019gauge}, which have important
applications in high energy density physics. Another noteworthy development
is the metriplectic particle-in-cell integrators for the Landau collision
operator \citep{Hirvijoki2017}. Recently, a field theory and a corresponding
6D structure-preserving geometric PIC algorithm were established for
low-frequency electrostatic perturbations in magnetized plasmas with
adiabatic electrons \citep{xiao2019field}. The long-term accuracy
and fidelity of the algorithm enabled the simulation study of electrostatic
drift wave turbulence and transport using a 6D PIC method with adiabatic
electrons.

The paper is organized as follows. In Sec.~\ref{Sec2} we develop
the explicit structure-preserving geometric PIC scheme in arbitrary
orthogonal coordinate systems, and the algorithm in a cylindrical
mesh is implemented in the\textsl{ SymPIC} code. In Sec.~\ref{Sec3},
the code is applied to carry out whole-device 6D kinetic PIC simulations
of tokamak physics.

\section{Explicit structure-preserving geometric PIC algorithm in curvilinear
orthogonal coordinate systems}

\label{Sec2}

\subsection{The basic principles of structure-preserving geometric PIC algorithm}

The procedure of designing a structure-preserving geometric PIC algorithm
starts from a field theory, or variational principle, for the system
of charged particles and electromagnetic field. Instead of discretizing
the corresponding Vlasov-Maxwell equations, the variational principle
is discretized.

For spatial discretization of the electromagnetic field, discrete
exterior calculus \citep{hirani2003discrete,desbrun2005discrete}
is adopted. As indicated by its name, a PIC algorithm contains two
components not found in other simulation methods: charge and current
deposition from particles to grid points, and field interpolation
from grid points to particles. Note that these two components are
independent from the field solver, e.g., the Yee-FDTD method, and
the particle pusher, e.g., the Boris algorithm. In conventional PIC
algorithms \citep{Dawson1983,hockney1988computer,birdsall1991plasma},
the function of charge and current deposition and the function of
field interpolation are implemented using intuitive techniques without
fundamental guiding principles other than the consistency requirement.
It was found \citep{squire4748,squire2012geometric,xiao2018structure}
that a systematic method to realize these two functions is to apply
the Whitney interpolation (differential) forms \citep{whitney1957geometric}
and their high-order generalizations \citep{xiao2015explicit} on
the discrete Lagrangian.

The application of Whitney forms is a key technology for achieving
the goal of preserving geometric structures and conservation laws.
It stipulates from first principles how charge and current deposition
and field interpolation should be performed \citep{squire4748,squire2012geometric,xiao2013variational,xiao2015explicit,xiao2015variational,he2015hamiltonian,qin2016canonical,he2016hamiltonian,kraus2017gempic,xiao2017local,xiao2018structure,xiao2019field}.
It also enabled the derivation of the discrete non-canonical symplectic
structure and Poisson bracket for the charged particle-electromagnetic
field system \citep{xiao2015explicit}. At the continuous limit when
the size of the space-time grid approaches zero, the discrete non-canonical
Poisson bracket reduces to the Morrison-Marsden-Weinstein (MMW) bracket
for the Vlasov-Maxwell equations \citep{morrison1980maxwell,marsden1982hamiltonian,weinstein1981comments,burby2017finite,Morrison2017}
(The MMW bracket was also independently discovered by Iwinski and
Turski \citep{Iwinski1976}). To derive the discrete Poisson bracket,
Whitney forms and their high-order generalizations are adopted for
the purpose of geometric spatial discretization of the field Lagrangian
density as an 1-form in the phase space of the charged particle-electromagnetic
field system, whose exterior derivative automatically generates a
discrete non-canonical symplectic structure and consequently a Poisson
bracket. As a different approach, He et al. \citep{he2016hamiltonian}
and Kraus et al. \citep{kraus2017gempic} used the method of finite
element exterior calculus to discretize the MMW bracket directly,
and verified explicitly that the discretized bracket satisfies the
Jacobi identity through lengthy calculations, in order for the discretized
bracket to be a legitimate Poisson bracket. If one chooses to discretize
the MMW bracket directly \citep{Perse2020,hirvijoki2020}, such verification
is necessary because there are other discretizations of the MMW bracket
which satisfy the numerical consistency requirement but not the Jacobi
identity. On the other hand, the discrete Poisson bracket in Ref.\,\citep{xiao2015explicit}
was not a discretization of the MMW bracket. It was derived from the
spatially discretized 1-form using Whitney forms and is naturally
endowed with a symplectic structure and Poisson bracket, and reduces
to the MMW bracket in the continuous limit. The advantage of this
discrete Poisson bracket in this respect demonstrates the power of
Whitney forms in the design of structure-preserving geometric PIC
algorithms.

To numerically integrate a Hamiltonian or Poisson system for the purpose
of studying multiscale physics, symplectic integrators are necessities.
Without an effective symplectic integrator, a symplectic or Poisson
structure is not beneficial in terms of providing a better algorithm
with long-term accuracy and fidelity. However, essentially all known
symplectic integrators are designed for canonical Hamiltonian systems
\citep{Devogelaere56,lee82,Ruth83,Feng85,Feng86,lee87,SanzSerna1988,veselov88,yoshida1990construction,Forest90,Channell90,Candy91,Tang93,Sanz-Serna94,Shang94,Feng95,Shang99,marsden2001discrete,Guo2002,Hairer02, Hong02,Shang2006,Feng10,zhang2016explicit,Tao2016},
except for recent investigations on non-canonical symplectic integrators
\citep{qin2008variational,qin2009variational,squire2012gauge,zhang2014canonicalization,Ellison2015,Burby2017,Kraus2017,Ellison2018,LelandThesis,he2017explicit,zhou2017explicit,xiao2019a,Shi2019,Xiao2020Slow}
for charged particle dynamics in a given electromagnetic field. Generic
symplectic integrators for non-canonical symplectic systems are not
known to exist. Fortunately, a high-order explicit symplectic integrator
for the MMW bracket was discovered \citep{he2015hamiltonian} using
a splitting technique in the Cartesian coordinate system. The validity
of this explicit non-canonical symplectic splitting method is built
upon the serendipitous one-way decoupling of the dynamic components
of the particle-field system in the Cartesian coordinate system. It's
immediately realized \citep{xiao2015explicit} that this explicit
non-canonical symplectic splitting applies to the discrete non-canonical
Poisson bracket without modification for the charged particle-electromagnetic
field system. This method was also adopted in Refs.\,\citep{he2016hamiltonian}
and \citep{kraus2017gempic} to integrate the discretized MMW bracket
using finite element exterior calculus. See also Refs.\,\citep{crouseilles2015}
and \citep{Qin15JCP} for early effort for developing symplectic splitting
method for the Vlasov-Maxwell system. It was recently proven \citep{xiao2018structure}
that the discrete non-canonical Poisson bracket \citep{xiao2015explicit}
and the explicit non-canonical symplectic splitting algorithm \citep{he2015hamiltonian}
can be equivalently formulated as a discrete field theory \citep{Qin2019}
using a Lagrangian discretized in both space and time.

The geometric discretization of the field theory for the system of
charged particles and electromagnetic field using Whitney forms \citep{whitney1957geometric,squire2012geometric,xiao2015explicit},
discrete exterior calculus \citep{hirani2003discrete,desbrun2005discrete},
and explicit non-canonical symplectic integration \citep{he2015hamiltonian,xiao2015explicit}
results in a PIC algorithm preserving many important physical symmetries
and conservation laws. In addition to the truncated infinitely dimensional
symplectic structure, the algorithms preserves exactly the local conservation
laws for charge \citep{squire4748,squire2012geometric,xiao2018structure}
and energy \citep{xiao2017local}. It was shown that the discrete
local charge conservation law is a natural consequence of the discrete
gauge symmetry admitted by the system \citep{squire4748,squire2012geometric,xiao2018structure,glasser2019b}.
The correspondence between discrete space-time translation symmetry
\citep{glasser2019,glasser2019a} and discrete local energy-momentum
conservation law has also been established for the Maxwell system
\citep{xiao2019Maxwell}.

As mentioned in Sec.\,\ref{sec:Introduction}, the goal of this section
is to extend the explicit structure-preserving geometric particle-in-cell
algorithm, especially the discrete non-canonical Poisson bracket \citep{xiao2015explicit}
and the explicit non-canonical symplectic splitting algorithm \citep{he2015hamiltonian},
from the Cartesian coordinate systems to curvilinear orthogonal coordinate
systems that are suitable for the tokamak geometry.

\subsection{Field theory of the particle-field system in curvilinear orthogonal
coordinate systems}

We start from the action integral of the system of charged particles
and electromagnetic field. In a curvilinear orthogonal coordinate
system $\left(x_{1},x_{2},x_{3}\right)$ with line element 
\begin{equation}
\rmd s^{2}=h_{1}\left(x_{1},x_{2},x_{3}\right)^{2}\rmd x_{1}^{2}+h_{2}\left(x_{1},x_{2},x_{3}\right)^{2}\rmd x_{2}^{2}+h_{3}\left(x_{1},x_{2},x_{3}\right)^{2}\rmd x_{3}^{2}\ ,
\end{equation}
the action integral of the system is 
\begin{eqnarray}
\calA[\bfx_{sp},\dot{\bfx}_{sp},\bfA,\phi] & = & \int\rmd t\sum_{s,p}\left(L_{sp}\left(m_{s},\bfv_{sp}\right)+q_{s}\left(\bfv_{sp}\cdot\bfA\left(\bfx_{sp},t\right)-\phi\left(\bfx_{sp},t\right)\right)\right)\nonumber \\
 &  & +\int\rmd V\rmd t\frac{1}{2}\left(\left(-\dot{\bfA}\left(\bfx,t\right)-\nabla\phi\left(\bfx,t\right)\right)^{2}-\left(\nabla\times\bfA\left(\bfx,t\right)\right)^{2}\right)~,
\end{eqnarray}
where $\bfx=[x_{1},x_{2},x_{3}]$ and 
$\rmd V=|h_{1}\left(\bfx\right)h_{2}\left(\bfx\right)h_{3}\left(\bfx\right)|\rmd x_{1}\rmd x_{2}\rmd x_{3}$.
In this coordinate system, the position and velocity of the $p$-th
particle of species $s$ are $\bfx_{sp}=[{x_{sp_{1}}},{x_{sp_{2}}},{x_{sp_{3}}}]$
and $\bfv_{sp}=[{\dot{x}_{sp_{1}}}h_{1}\left(\bfx_{sp}\right),{\dot{x}_{sp_{2}}}h_{2}\left(\bfx_{sp}\right),{\dot{x}_{sp_{3}}}h_{3}\left(\bfx_{sp}\right)]$.
$L_{sp}$ is the free Lagrangian for the $p$-th particle of species
$s$. For the non-relativistic case $L_{sp}=m_{s}|\bfv_{sp}|^{2}/2$,
and for the relativistic case $L_{sp}=-m_{s}\sqrt{1-|\bfv_{sp}|^{2}}$.
Here, we set both permittivity $\epsilon_{0}$ and permeability $\mu_{0}$
in the vacuum to 1 to simplify the notation. The evolution of this
system is governed by the variational principle, 
\begin{eqnarray}
\frac{\delta\calA}{\delta\bfA} & = & 0~,\label{EqnDADA}\\
\frac{\delta\calA}{\delta\phi} & = & 0~,\label{EqnDADP}\\
\frac{\delta\calA}{\delta\bfx_{sp}} & = & 0~.\label{EqnDADX}
\end{eqnarray}
It can be verified that \EQ{EqnDADA} and \EQ{EqnDADP} are the
Maxwell equations, and \EQ{EqnDADX} is Newton's equation for the
$p$-th particle of species $s$.

\subsection{Construction of the algorithm as a discrete field theory\label{Sec2Construction}}

According to the general principle of structure-preserving geometric
algorithm, the first step is to discretize the field theory \citep{squire4748,squire2012geometric,xiao2015explicit,xiao2018structure,xiao2019field}.
For the electromagnetic field, we use the technique of Whitney forms
and discrete manifold \citep{whitney1957geometric,hirani2003discrete}.
For example, in a tetrahedron mesh the magnetic field $\bfB$ as a
2-form field can be discretized on a 2-simplex $\sigma_{2}$ (a side
of a 3D tetrahedron mesh) as $\int_{\sigma_{2}}\bfB\cdot\rmd\mathbf{S}$,
where $\mathbf{S}$ is the unit normal vector of $\sigma_{2}$. Note
that $\sigma_{2}$ is a common side of two tetrahedron cells. In a
mesh constructed using a curvilinear orthogonal coordinate system,
which will be referred to as a Curvilinear Orthogonal Mesh (COM),
such discretization can be also performed. Let 
\begin{eqnarray}
\phi_{i,j,k,l} & = & \phi\left(x_{J,l}^{4}\right)~,\\
\bfA_{i,j,k,l} & = & \left[\begin{array}{c}
A_{x_{1}}\left(x_{J,l}^{4}\right)h_{1}\left(x_{J,l}^{4}\right),\\
A_{x_{2}}\left(x_{J,l}^{4}\right)h_{2}\left(x_{J,l}^{4}\right),\\
A_{x_{3}}\left(x_{J,l}^{4}\right)h_{3}\left(x_{J,l}^{4}\right)
\end{array}\right]^{T}~,\\
\bfB_{i,j,k,l} & = & \left[\begin{array}{c}
B_{x_{1}}\left(x_{J,l}^{4}\right)h_{2}\left(x_{J,l}^{4}\right)h_{3}\left(x_{J,l}^{4}\right),\\
B_{x_{2}}\left(x_{J,l}^{4}\right)h_{3}\left(x_{J,l}^{4}\right)h_{1}\left(x_{J,l}^{4}\right),\\
B_{x_{3}}\left(x_{J,l}^{4}\right)h_{1}\left(x_{J,l}^{4}\right)h_{2}\left(x_{J,l}^{4}\right)
\end{array}\right]^{T}~,\\
\rho_{i,j,k,l} & = & \rho\left(x_{J,l}^{4}\right)h_{1}\left(x_{J,l}^{4}\right)h_{2}\left(x_{J,l}^{4}\right)h_{3}\left(x_{J,l}^{4}\right)~,
\end{eqnarray}
where $\left(x_{J,l}^{4}\right)$ denotes $\left(i\Delta x_{1},j\Delta x_{2},k\Delta x_{3},l\Delta t\right)$,
$\phi_{i,j,k,l},\bfA_{i,j,k,l},\bfB_{i,j,k,l}$ and $\rho_{i,j,k,l}$
are discrete 0-, 1-, 2- and 3-forms, respectively. In this discretization,
the discrete gradient, curl and divergence operators can simply be
finite difference operators. In the present study, the following difference
operators are adopted, 
\begin{eqnarray}
\left({\nabla_{\mathrm{d}}}\phi\right)_{i,j,k} & = & [\phi_{i+1,j,k}-\phi_{i,j,k},\phi_{i,j+1,k}-\phi_{i,j,k},\phi_{i,j,k+1}-\phi_{i,j,k}]~,\label{EqnDEFGRADD}\\
\left(\CURLD\bfA\right)_{i,j,k} & = & \left[\begin{array}{c}
\left({A_{x_{3}}}_{i,j+1,k}-{A_{x_{3}}}_{i,j,k}\right)-\left({A_{x_{2}}}_{i,j,k+1}-{A_{x_{2}}}_{i,j,k}\right)\\
\left({A_{x_{1}}}_{i,j,k+1}-{A_{x_{1}}}_{i,j,k}\right)-\left({A_{x_{3}}}_{i+1,j,k}-{A_{x_{3}}}_{i,j,k}\right)\\
\left({A_{x_{2}}}_{i+1,j,k}-{A_{x_{2}}}_{i,j,k}\right)-\left({A_{x_{1}}}_{i,j+1,k}-{A_{x_{1}}}_{i,j,k}\right)
\end{array}\right]^{T}~,\label{DEFCURLD}\\
\left({\DIVD}\bfB\right)_{i,j,k} & = & \left({B_{x_{1}}}_{i+1,j,k}-{B_{x_{1}}}_{i,j,k}\right)+\left({B_{x_{2}}}_{i,j+1,k}-{B_{x_{2}}}_{i,j,k}\right)+\nonumber \\
 &  & \left({B_{z}}_{i,j,k+1}-{B_{z}}_{i,j,k}\right)~.\label{EqnDEFDIVD}
\end{eqnarray}
To discretize the particle-field interaction while preserving geometric
structures and symmetries, Whitney interpolating forms \citep{whitney1957geometric}
and their high-order generalizations \citep{xiao2015explicit} are
used. Akin to previous results on Whitney interpolating forms in a
cubic mesh \citep{xiao2015explicit}, interpolating forms $\WOZERO{\bfx},\WOONE{\bfx},\WOTWO{\bfx}$
and $\WOTHREE{\bfx}$ for 0-, 1-, 2- and 3-forms can be constructed
on a COM as follows, 
\begin{eqnarray}
\WOZERO{\bfx} & = & \WZERO{\bfx}~,\\
\WOONE{\bfx} & = & \WONE{\bfx}/[h_{1}\left(\bfx\right),h_{2}\left(\bfx\right),h_{3}\left(\bfx\right)]~,\\
\WOTWO{\bfx} & = & \WTWO{\bfx}/[h_{2}\left(\bfx\right)h_{3}\left(\bfx\right),h_{3}\left(\bfx\right)h_{1}\left(\bfx\right),h_{1}\left(\bfx\right)h_{2}\left(\bfx\right)]~,\\
\WOTHREE{\bfx} & = & \WTHREE{\bfx}/h_{1}\left(\bfx\right)h_{2}\left(\bfx\right)h_{3}\left(\bfx\right)~,
\end{eqnarray}
where $W_{\sigma iJ}\left(\bfx\right),0\leq i\leq3$ are the interpolating
forms in a cubic mesh defined in Ref.~\citep{xiao2015explicit} and
the quotient (product) of vectors means component-wise quotient (product),
i.e., 
\begin{eqnarray}
[A_{x_{1}},A_{x_{2}},A_{x_{3}}]/[B_{x_{1}},B_{x_{2}},B_{x_{3}}] & = & [\frac{A_{x_{1}}}{B_{x_{1}}},\frac{A_{x_{2}}}{B_{x_{2}}},\frac{A_{x_{3}}}{B_{x_{3}}}]~.
\end{eqnarray}

With these operators and interpolating forms, we discretize the action
integral as 
\begin{eqnarray}
\calA_{d} & = & \sum_{s,p,l}\left(L_{sp}\left(m_{s},\bfv_{sp,l}\right)+q_{s}\left(\overline{\bfv_{sp}\cdot\bfA}\left(\bfx_{sp,l},\bfx_{sp,l+1},l\right)-\phi\left(\bfx_{sp,l+1},l+1\right)\right)\right)+\nonumber \\
 &  & \frac{\bfhc\left(\bfx_{J}\right)}{2}\sum_{J}\left(\left(-\frac{\bfA_{J,l+1}-\bfA_{J,l}}{\Delta t\bfha\left(\bfx_{J}\right)}-\frac{\left(\nabla_{d}\phi\right)_{J,l+1}}{\bfha\left(\bfx_{J}\right)}\right)^{2}-\left(\frac{\left(\CURLD\bfA\right)_{K,l}}{\bfhb\left(\bfx_{K}\right)}\right)^{2}\right)~,
\end{eqnarray}
where $l$ is the index for the temporal grid, $J$ is the index vector
for the spatial grid, and 
\begin{eqnarray}
\bfha\left(\bfx\right) & = & [h_{1}\left(\bfx\right),h_{2}\left(\bfx\right),h_{3}\left(\bfx\right)]~,\\
\bfhb\left(\bfx\right) & = & [h_{2}\left(\bfx\right)h_{3}\left(\bfx\right),h_{3}\left(\bfx\right)h_{1}\left(\bfx\right),h_{1}\left(\bfx\right)h_{2}\left(\bfx\right)]~,\\
\bfhc\left(\bfx\right) & = & h_{1}\left(\bfx\right)h_{2}\left(\bfx\right)h_{3}\left(\bfx\right)~,\\
\bfv_{sp,l} & = & \frac{\bfx_{sp,l+1}-\bfx_{sp,l}}{\Delta t}\bfha\left(\frac{\bfx_{sp,l+1}+\bfx_{sp,l}}{2}\right)~,\label{EqnDV}\\
\phi\left(\bfx,l\right) & = & \sum_{J}\phi_{J,l}\WZERO{\bfx}~,\\
\overline{\bfv_{sp}\cdot\bfA}\left(\bfx_{sp,l},\bfx_{sp,l+1},l\right) & = & \int_{0}^{1}\rmd\tau\left[\begin{array}{c}
\frac{x_{1,sp,l+1}-x_{1,sp,l}}{\Delta t}h_{1}\left(x_{1,sp,l}+\tau\left(x_{1,sp,l+1}-x_{1,sp,l},x_{2,sp,l},x_{3,sp,l}\right)\right),\\
\frac{x_{2,sp,l+1}-x_{2,sp,l}}{\Delta t}h_{2}\left(x_{1,sp,l+1},x_{2,sp,l}+\tau\left(x_{2,sp,l+1}-x_{2,sp,l},x_{3,sp,l}\right)\right),\\
\frac{x_{3,sp,l+1}-x_{3,sp,l}}{\Delta t}h_{3}\left(x_{1,sp,l+1},x_{2,sp,l+1},x_{3,sp,l}+\tau\left(x_{3,sp,l+1}-x_{3,sp,l}\right)\right)
\end{array}\right]\nonumber \\
 &  & \cdot\left[\begin{array}{l}
\tilde{A}_{x_{1},l}\left(x_{1,sp,l}+\tau\left(x_{1,sp,l+1}-x_{1,sp,l}\right),x_{2,sp,l},x_{3,sp,l}\right),\\
\tilde{A}_{x_{2},l}\left(x_{1,sp,l+1},x_{2,sp,l}+\tau\left(x_{2,sp,l+1}-x_{2,sp,l}\right),x_{3,sp,l}\right),\\
\tilde{A}_{x_{3},l}\left(x_{1,sp,l+1},x_{2,sp,l+1},x_{3,sp,l}+\tau\left(x_{3,sp,l+1}-x_{3,sp,l}\right)\right)
\end{array}\right]\\
 & = & \int_{0}^{1}\rmd\tau\left[\frac{x_{1,sp,l+1}-x_{1,sp,l}}{\Delta t},\frac{x_{2,sp,l+1}-x_{2,sp,l}}{\Delta t},\frac{x_{3,sp,l+1}-x_{3,sp,l}}{\Delta t}\right]\nonumber \\
 &  & \cdot\left[\begin{array}{l}
A_{x_{1},l}\left(x_{1,sp,l}+\tau\left(x_{1,sp,l+1}-x_{1,sp,l}\right),x_{2,sp,l},x_{3,sp,l}\right),\\
A_{x_{2},l}\left(x_{1,sp,l+1},x_{2,sp,l}+\tau\left(x_{2,sp,l+1}-x_{2,sp,l}\right),x_{3,sp,l}\right),\\
A_{x_{3},l}\left(x_{1,sp,l+1},x_{2,sp,l+1},x_{3,sp,l}+\tau\left(x_{3,sp,l+1}-x_{3,sp,l}\right)\right)
\end{array}\right]~,\label{EqnACTINT}
\end{eqnarray}
\begin{eqnarray}
\left[\begin{array}{c}
\tilde{A}_{x_{1},l}\left(\bfx\right),\\
\tilde{A}_{x_{2},l}\left(\bfx\right),\\
\tilde{A}_{x_{3},l}\left(\bfx\right)
\end{array}\right] & \equiv & \sum_{J}\bfA_{J,l}\WOONE{\bfx}~,\\
\left[\begin{array}{c}
A_{x_{1},l}\left(\bfx\right),\\
A_{x_{2},l}\left(\bfx\right),\\
A_{x_{3},l}\left(\bfx\right)
\end{array}\right] & \equiv & \sum_{J}\bfA_{J,l}\WONE{\bfx}~.
\end{eqnarray}
Finally, the time advance rule is given by the variation of the action
integral with respect to the discrete fields, 
\begin{eqnarray}
\frac{\partial\calA_{d}}{\partial\bfx_{sp,l}} & = & 0~,\label{EqnDLF}\\
\frac{\partial\calA_{d}}{\partial\bfA_{J,l}} & = & 0~,\label{EqnDLA}\\
\frac{\partial\calA_{d}}{\partial\phi_{J,l}} & = & 0~.\label{EqnDLP}
\end{eqnarray}

From \EQ{EqnDLF}, the variation with respect to $\bfx_{sp,l}$
leads to 
\begin{eqnarray}
\frac{\partial}{\partial\bfx_{l}}\left(L_{sp}\left(m_{s},\bfv_{sp,l-1}\right)+L_{sp}\left(m_{s},\bfv_{sp,l}\right)\right) & = & q_{s}\left(\bfE_{l}\left(\bfx_{sp,l}\right)+\overline{\bfv_{sp}\times\bfB}\left(l-1,l,l+1\right)\right)~,
\end{eqnarray}
where 
\begin{eqnarray}
\bfE_{l}\left(\bfx\right) & = & \sum_{J}\bfE_{J,l}\WONE{\bfx}~,\\
\bfE_{J,l} & = & -\frac{\bfA_{J,l}-\bfA_{J,l-1}}{\Delta t}~,\\
\overline{\bfv_{sp}\times\bfB}\left(l-1,l,l+1\right) & = & \frac{1}{\Delta t}\left[\begin{array}{l}
\int_{x_{2,sp,l-1}}^{x_{2,sp,l}}\rmd x_{2}B_{x_{3},l-1}\left(x_{1,sp,l},x_{2},x_{3,sp,l-1}\right),\\
\int_{x_{3,sp,l-1}}^{x_{3,sp,l}}\rmd x_{3}B_{x_{1},l-1}\left(x_{1,sp,l},x_{2,sp,l},x_{3}\right),\\
\int_{x_{1,sp,l}}^{x_{1,sp,l+1}}\rmd x_{1}B_{x_{2},l}\left(x_{1},x_{2,sp,l},x_{3,sp,l}\right)
\end{array}\right]\nonumber \\
 &  & -\frac{1}{\Delta t}\left[\begin{array}{l}
\int_{x_{3,sp,l-1}}^{x_{3,sp,l}}\rmd x_{3}B_{x_{2},l-1}\left(x_{1,sp,l},x_{2,sp,l},x_{3}\right),\\
\int_{x_{1,sp,l}}^{x_{1,sp,l+1}}\rmd x_{1}B_{x_{3},l}\left(x_{1},x_{2,sp,l},x_{3,sp,l}\right),\\
\int_{x_{2,sp,l}}^{x_{2,sp,l+1}}\rmd x_{2}B_{x_{1},l}\left(x_{1,sp,l+1},x_{2},x_{3,sp,l}\right)
\end{array}\right]~,\\
\left[\begin{array}{c}
B_{x_{1},l}\left(\bfx\right),\\
B_{x_{2},l}\left(\bfx\right),\\
B_{x_{3},l}\left(\bfx\right)
\end{array}\right] & = & \sum_{J}\bfB_{J,l}\WTWO{\bfx}~,\\
\bfB_{K,l} & = & \sum_{J}\CURLD_{K,J}\bfA_{J,l}~.\label{EqnDEFB}
\end{eqnarray}
To reduce simulation noise, we use 2nd-order Whitney forms for field
interpolation. The concept of 2nd-order Whitney forms and their constructions
were systematically developed in \REF{xiao2015explicit}. In general,
only piece-wise polynomials are used to construct high-order Whitney
forms, and it is straightforward to calculate integrals along particles'
trajectories. These integrals can be calculated explicitly even with
more complex Whitney interpolating forms, because these interpolating
forms contain derivatives that are easy to integrate.

From \EQ{EqnDLA}, the discrete Ampere's law is 
\[
\frac{1}{\bfha\left(\bfx_{J}\right)\bfha\left(\bfx_{J}\right)}\frac{\bfE_{J,l+1}-\bfE_{J,l}}{\Delta t}\bfhc\left(\bfx_{J}\right)=\sum_{K}\CURLDP_{J,K}\frac{\bfhc\left(\bfx_{J}\right)}{\bfhb\left(\bfx_{K}\right)\bfhb\left(\bfx_{K}\right)}\bfB_{K,l}-\bfJ_{J,l}~,
\]
where 
\begin{eqnarray}
\bfJ_{J,l} & = & \frac{1}{\Delta t}\sum_{s,p}q_{s}\int_{C_{sp,l,l+1}}\WONE{\bfx}\rmd\bfx~,
\end{eqnarray}
and the integral path $C_{sp,l,l+1}$ is defined as a zigzag path
from $\bfx_{sp,l}$ to $\bfx_{sp,l+1}$, i.e., 
\begin{eqnarray}
C_{sp,l,l+1} & = & \left\{ \left(x_{1,sp,l}+\tau\left(x_{1,sp,l+1}-x_{1,sp,l}\right),x_{2,sp,l},x_{3,sp,l}\right)|\tau\in\left[0,1\right)\right\} \bigcup\nonumber \\
 &  & \left\{ \left(x_{1,sp,l+1},x_{2,sp,l}+\tau\left(x_{2,sp,l+1}-x_{2,sp,l}\right),x_{3,sp,l}\right)|\tau\in\left[0,1\right)\right\} \bigcup\nonumber \\
 &  & \left\{ \left(x_{1,sp,l+1},x_{2,sp,l+1},x_{3,sp,l}+\tau\left(x_{3,sp,l+1}-x_{3,sp,l}\right)\right)|\tau\in\left[0,1\right)\right\} ~.
\end{eqnarray}
Using the definition of $\bfB_{K,l}$, i.e., \EQ{EqnDEFB}, we can
obtain its discrete time evolution, 
\begin{eqnarray}
\frac{\bfB_{K,l}-\bfB_{K,l-1}}{\Delta t} & = & -\sum_{J}\CURLD_{K,J}\bfE_{J,l}~,
\end{eqnarray}
which is the discrete version of Faraday's law.

Generally the above scheme is implicit. However, if particles are
non-relativistic and the line element vector $\mathbf{h}$ of a curvilinear
orthogonal coordinate system satisfies the following condition, 
\begin{eqnarray}
\frac{\partial h_{1}}{\partial x_{1}}=\frac{\partial h_{2}}{\partial x_{2}}=\frac{\partial h_{3}}{\partial x_{3}}=0~,\label{EqnH123}
\end{eqnarray}
then high-order explicit schemes exist in the COM constructed using
this coordinate system. The cylindrical coordinate is such a case.
In Sec.\,\ref{Sec2cylindrical} we derive the 2nd-order explicit
scheme for the non-relativistic Vlasov-Maxwell system in the cylindrical
mesh.

\subsection{Poisson bracket and its splitting algorithm in a curvilinear orthogonal
mesh\label{Sec2XQ-bracket}}

For the structure-preserving geometric PIC algorithm presented in
Sec.\,\ref{Sec2Construction}, there exists a corresponding discrete
Poisson bracket. When particles are non-relativistic and condition
(\ref{EqnH123}) is satisfied, an associate splitting algorithm, which
is explicit and symplectic, can be constructed. The algorithm is similar
to and generalizes the explicit non-canonical symplectic splitting
algorithm in the Cartesian coordinate system designed in Ref.\,\citep{he2015hamiltonian}.
Since the algorithm formulated in Sec.\,\ref{Sec2Construction} is
independent from these constructions, we only list the results here
without detailed derivations.

In a COM built on a curvilinear orthogonal coordinate system, we can
choose to discretize the space only using the same method described
above to obtain 
\begin{align}
L_{sd} & =\sum_{s,p}\left(L_{sp}\left(m_{s},\dot{\bfx}_{sp}\right)+q_{s}\dot{\bfx}_{sp}\cdot\sum_{J}\bfA_{J}\WONE{\bfx_{sp}}\right)+\nonumber \\
 & \frac{1}{2}\sum_{J}\left(\left(\frac{-\dot{\bfA}_{J}}{\bfha\left(\bfx_{J}\right)}\right)^{2}-\left(\frac{\left(\CURLD\bfA\right)_{J,l}}{\bfhb\left(\bfx_{J}\right)}\right)^{2}\right)\bfhc\left(\bfx_{J}\right)~,\label{EqAdSpace}
\end{align}
where the temporal gauge ($\phi=0$) has been adopted. Following the
procedure in Ref.~\citep{xiao2015explicit}, a non-canonical symplectic
structure can be constructed from this Lagrangian, and the associated
discrete Poisson bracket is 
\begin{eqnarray}
\left\{ F,G\right\}  & = & \sum_{J}\left(\frac{\partial F}{\partial\bfE_{J}}\cdot\frac{\diag\left(\bfha\left(\bfx_{J}\right)^{2}\right)}{\bfhc\left(\bfx_{J}\right)}\cdot\sum_{K}\frac{\partial G}{\partial\bfB_{K}}\CURLD_{KJ}-\right.\nonumber \\
 &  & \left.\sum_{K}\frac{\partial F}{\partial\bfB_{K}}\CURLD_{KJ}\cdot\frac{\diag\left(\bfha\left(\bfx_{J}\right)^{2}\right)}{\bfhc\left(\bfx_{J}\right)}\cdot\frac{\partial G}{\partial\bfE_{J}}\right)+\nonumber \\
 &  & \sum_{s,p}\frac{1}{m_{s}}\sum_{i=1}^{3}\frac{1}{h_{i}\left(\bfx_{sp}\right)^{2}}\left(\frac{\partial F}{\partial x_{i,sp}}\cdot\frac{\partial G}{\partial\dot{x}_{i,sp}}-\frac{\partial F}{\partial\dot{x}_{i,sp}}\cdot\frac{\partial G}{\partial x_{i,sp}}\right)+\sum_{s,p}\sum_{J}\sum_{i=1}^{3}\nonumber \\
 &  & \frac{\nbfha\left(\bfx_{J}\right)^{2}q_{s}}{m_{s}h_{i}\left(\bfx_{sp}\right)^{2}\nbfhc\left(\bfx_{J}\right)}\left(\frac{\partial F}{\partial\dot{x}_{i,sp}}W_{\sigma_{1J}}\left(x_{i,sp}\right)\frac{\partial G}{\partial E_{x_{i},J}}-\frac{\partial G}{\partial\dot{x}_{i,sp}}W_{\sigma_{1J}}\left(x_{i,sp}\right)\frac{\partial F}{\partial E_{x_{i},J}}\right)\nonumber \\
 &  & +\sum_{s}\sum_{K}\frac{1}{m_{s}^{2}}\left(\diag\left(\frac{1}{\bfha\left(\bfx_{sp}\right)^{2}}\right)\cdot\frac{\partial F}{\partial\dot{\bfx}_{sp}}\right)\cdot\nonumber \\
 &  & \Bigg(\left(\nabla_{\bfx_{sp}}\times\left(m_{s}\diag\left(\bfha\left(\bfx_{sp}\right)^{2}\right)\cdot\dot{\bfx}_{sp}\right)+q_{s}W_{\sigma_{2K}}\left(\bfx_{sp}\right)\bfB_{K}\right)\nonumber \\
 &  & \times\left(\frac{\partial G}{\partial\dot{\bfx}_{sp}}\cdot\diag\left(\frac{1}{\bfha\left(\bfx_{sp}\right)^{2}}\right)\right)\Bigg)~,\label{XQbracket}
\end{eqnarray}
and the corresponding Hamiltonian is 
\begin{eqnarray}
H_{sd} & = & \frac{1}{2}\sum_{J}\left(\left|\frac{\bfE_{J}}{\bfha\left(\bfx_{J}\right)}\right|^{2}+\left|\frac{\bfB_{J,l}}{\bfhb\left(\bfx_{J}\right)}\right|^{2}\right)\bfhc\left(\bfx_{J}\right)+\sum_{sp}\sum_{i}\frac{m_{s}}{2}\dot{x}_{i,sp}^{2}h_{i}\left(\bfx_{sp}\right)^{2}~.
\end{eqnarray}
Here particles are assumed to be non-relativistic and $\bfE_{J},\bfB_{K}$
are spatially discretized electromagnetic fields defined as

\begin{eqnarray}
\bfE_{J} & = & -\dot{\bfA}_{J}~,\\
\bfB_{K} & = & \sum_{J}\CURLD_{KJ}\bfA_{J}~.
\end{eqnarray}
The Poisson bracket given by Eq.\,(\ref{XQbracket}) generalizes
the previous Cartesian version \citep{xiao2015explicit} to arbitrary
curvilinear orthogonal meshes. It automatically satisfies the Jacobi
identity because it is derived from a Lagrangian 1-form. See Ref.\ \citep{xiao2015explicit}
for detailed geometric constructions.

The dynamics equation, i.e., the Hamiltonian equation, is

\begin{equation}
\dot{F}=\{F,H_{sd}\}~,
\end{equation}
where 
\begin{equation}
F=[\bfE_{J},\bfB_{J},\bfx_{sp},\dot{\bfx}_{sp}]~.
\end{equation}
To build an explicit symplectic algorithm, we adopt the splitting
method \citep{he2015hamiltonian}. The Hamiltonian $H_{sd}$ is divided
into 5 parts, 
\begin{eqnarray}
H_{sd} & = & H_{E}+H_{B}+H_{1}+H_{2}+H_{3}~,\\
H_{E} & = & \frac{1}{2}\sum_{J}\left|\frac{\bfE_{J}}{\bfha\left(\bfx_{J}\right)}\right|^{2}\bfhc\left(\bfx_{J}\right)~,\\
H_{B} & = & \frac{1}{2}\sum_{J}\left|\frac{\bfB_{J,l}}{\bfhb\left(\bfx_{J}\right)}\right|^{2}\bfhc\left(\bfx_{J}\right)~,\\
H_{i} & = & \sum_{sp}\frac{m_{s}}{2}\dot{x}_{i,sp}^{2}h_{i}\left(\bfx_{sp}\right)^{2}\quad\textrm{for }i\textrm{ in}~\left\{ 1,2,3\right\} ~.
\end{eqnarray}
Each part defines a sub-system with the same Poisson bracket (\ref{XQbracket}).
It turns out that when condition (\ref{EqnH123}) is satisfied the
exact solution of each subsystem can be written down in a closed form,
and explicit high-order symplectic algorithms for the entire system
can be constructed by compositions using the exact solutions of the
sub-systems. For $H_{E}$ and $H_{B}$, the corresponding Hamiltonian
equations are 
\begin{eqnarray}
\dot{F} & = & \left\{ F,H_{E}\right\} ~,\\
\dot{F} & = & \left\{ F,H_{B}\right\} ~,
\end{eqnarray}
i.e., 
\begin{eqnarray}
\left\{ \begin{array}{ccl}
\dot{\bfE}_{J} & = & 0~,\\
\dot{\bfB}_{K} & = & -\sum_{J}\CURLD_{KJ}\bfE_{J}~,\\
\dot{\bfx}_{sp} & = & 0~,\\
\ddot{\bfx}_{sp} & = & \frac{q_{s}}{m_{s}}\diag\left(\frac{1}{\bfha\left(\bfx_{sp}\right)^{2}}\right)\cdot\sum_{J}W_{\sigma_{1J}}\left(\bfx_{sp}\right)\bfE_{J}~,
\end{array}\right.
\end{eqnarray}
and 
\begin{eqnarray}
\left\{ \begin{array}{ccl}
\dot{\bfE}_{J} & = & \frac{\diag\left(\bfha\left(\bfx_{J}\right)^{2}\right)}{\bfhc\left(\bfx_{J}\right)}\cdot\sum_{K}\bfhc\left(\bfx_{K}\right)\diag\left(\frac{1}{\bfhb\left(\bfx_{K}\right)^{2}}\right)\cdot\CURLD_{KJ}\bfB_{K}~,\\
\dot{\bfB}_{K} & = & 0~,\\
\dot{\bfx}_{sp} & = & 0~,\\
\ddot{\bfx}_{sp} & = & 0\ .
\end{array}\right.
\end{eqnarray}
Their analytical solutions are 
\begin{eqnarray}
\Theta_{E}:\left\{ \begin{array}{ccl}
\bfE_{J}\left(t+\Delta t\right) & = & \bfE_{J}\left(t\right)~,\\
\bfB_{K}\left(t+\Delta t\right) & = & \bfB_{K}\left(t\right)-\Delta t\sum_{J}\CURLD_{KJ}\bfE_{J}(t)~,\\
\bfx_{sp}\left(t+\Delta t\right) & = & \bfx_{s}\left(t\right)~,\\
\dot{\bfx}_{sp}\left(t+\Delta t\right) & = & \dot{\bfx}_{s}\left(t\right)+\Delta t\frac{q_{s}}{m_{s}}\diag\left(\frac{1}{\bfha\left(\bfx_{sp}\right)^{2}}\right)\cdot\sum_{J}W_{\sigma_{1J}}\left(\bfx_{sp}(t)\right)\bfE_{J}(t)~,
\end{array}\right.
\end{eqnarray}
\begin{eqnarray}
\Theta_{B}:\left\{ \begin{array}{ccl}
\bfE_{J}\left(t+\Delta t\right) & = & \bfE_{J}\left(t\right)+\Delta t\frac{\diag\left(\bfha\left(\bfx_{J}\right)^{2}\right)}{\bfhc\left(\bfx_{J}\right)}\cdot\sum_{K}\CURLD_{KJ}\bfhc\left(\bfx_{K}\right)\diag\left(\frac{1}{\bfhb\left(\bfx_{K}\right)^{2}}\right)\cdot\bfB_{K}(t)~,\\
\bfB_{K}\left(t+\Delta t\right) & = & \bfB_{K}\left(t\right)~,\\
\bfx_{sp}\left(t+\Delta t\right) & = & \bfx_{sp}\left(t\right)~,\\
\dot{\bfx}_{sp}\left(t+\Delta t\right) & = & \dot{\bfx}_{sp}\left(t\right)~.
\end{array}\right.
\end{eqnarray}
For $H_{1}$, the dynamic equation is $\dot{F}=\{F,H_{1}\}$, or more
specifically, 
\begin{eqnarray}
\left\{ \begin{array}{ccl}
\dot{\bfE}_{J} & = & -\sum_{sp}\frac{q_{s}}{\bfhc\left(\bfx_{J}\right)}\diag\left(\bfha\left(\bfx_{J}\right)^{2}\right)\cdot\dot{x}_{1,sp}\bfe_{1}W_{\sigma_{1J}}\left(\bfx_{sp}\right)~,\\
\dot{\bfB}_{K} & = & 0~,\\
\dot{\bfx}_{sp} & = & \dot{x}_{1,sp}\bfe_{1}~,\\
\ddot{\bfx}_{sp} & =- & \diag\left(\frac{1}{2\bfha\left(\bfx_{sp}\right)^{2}}\right)\cdot\nabla_{\bfx_{sp}}h_{1}\left(\bfx_{sp}\right)^{2}\dot{x}_{1,sp}^{2}+\newline\\
 &  & \diag\left(\frac{1}{\bfha\left(\bfx_{sp}\right)^{2}}\right)\dot{x}_{1,sp}\bfe_{1}\times\left(\nabla_{\bfx_{sp}}\times\left(\diag\left(\bfha\left(\bfx_{sp}\right)^{2}\cdot\dot{\bfx}_{sp}\right)\right)\right)+\\
 &  & \diag\left(\frac{q_{s}}{\bfha\left(\bfx_{sp}\right)^{2}m_{s}}\right)\dot{x}_{1,sp}\bfe_{1}\times\sum_{K}W_{\sigma_{2K}}\left(\bfx_{s}\right)\bfB_{K}~.
\end{array}\label{EqEVOX}\right.
\end{eqnarray}
Because the equation for $x_{1,sp}$ contains both $\dot{x}_{1,sp}$
and $\ddot{x}_{1,sp}$explicitly, \EQ{EqEVOX} is difficult to solve
in general. However, when $\partial h_{1}\left(\bfx\right)/\partial x_{1}=0$,
i.e., 
\begin{eqnarray}
h_{1}\left(\bfx\right)=h_{1}\left(x_{2},x_{3}\right),
\end{eqnarray}
the dynamics equation for particles in \EQ{EqEVOX} reduces to 
\begin{eqnarray}
\left\{ \begin{array}{rcl}
\dot{x}_{1,sp} & = & \dot{x}_{1,sp}~,\\
\dot{x}_{2,sp} & = & 0~,\\
\dot{x}_{3,sp} & = & 0~,\\
\ddot{x}_{1,sp}h_{1}\left(\bfx_{sp}\right)^{2} & = & 0~,\\
\frac{\rmd}{\rmd t}\left(\dot{x}_{2,sp}h_{2}\left(\bfx_{sp}\right)^{2}\right) & = & 2\dot{x}_{1,sp}^{2}h_{1}\left(\bfx_{sp}\right)\frac{\partial h_{1}\left(\bfx_{sp}\right)}{\partial x_{2,sp}}-\frac{q_{s}}{m_{s}}\dot{x}_{1,sp}\sum_{K}W_{\sigma_{2K},x_{3}}\left(\bfx_{sp}\right)B_{x_{3},K}\\
 & = & \tilde{F}_{x_{2}}\left(\bfx_{sp},\dot{x}_{1,sp}\right)~,\\
\frac{\rmd}{\rmd t}\left(\dot{x}_{3,sp}h_{3}\left(\bfx_{sp}\right)^{2}\right) & = & 2\dot{x}_{1,sp}^{2}h_{1}\left(\bfx_{sp}\right)\frac{\partial h_{1}\left(\bfx_{sp}\right)}{\partial x_{3,sp}}+\frac{q_{s}}{m_{s}}\dot{x}_{1,sp}\sum_{K}W_{\sigma_{2K},x_{2}}\left(\bfx_{sp}\right)B_{x_{2},K}\\
 & = & \tilde{F}_{x_{3}}\left(\bfx_{sp},\dot{x}_{1,sp}\right)~.
\end{array}\right.
\end{eqnarray}
In this case, \EQ{EqEVOX} admits an analytical solution, 
\begin{eqnarray}
\Theta_{1}\left(\Delta t\right):\left\{ \begin{array}{ccl}
{\bfE}_{J}\left(t+\Delta t\right) & = & -\sum_{sp}\frac{q_{s}}{\bfhc\left(\bfx_{J}\right)}\diag\left(\bfha\left(\bfx_{J}\right)^{2}\right)\cdot\int_{0}^{\Delta t}\rmd t'\dot{x}_{1,sp}\bfe_{1}W_{\sigma_{1J}}\left(\bfx_{sp}+\bfe_{1}\dot{x}_{1,sp}t'\right)~,\\
\bfB_{K}\left(t+\Delta t\right) & = & \bfB_{K}\left(t\right)~,\\
x_{1,sp}\left(t+\Delta t\right) & = & x_{1,sp}\left(t\right)+\dot{x}_{1,sp}\left(t\right)\Delta t~,\\
x_{2,sp}\left(t+\Delta t\right) & = & x_{2,sp}\left(t\right)~,\\
x_{3,sp}\left(t+\Delta t\right) & = & x_{3,sp}\left(t\right)~,\\
\dot{x}_{1,sp}\left(t+\Delta t\right) & = & \dot{x}_{1,sp}\left(t\right)~,\\
\dot{x}_{2,sp}\left(t+\Delta t\right) & = & \frac{h_{2}\left(\bfx_{sp}\left(t\right)\right)^{2}}{h_{2}\left(\bfx_{sp}\left(t\right)+\bfe_{1}\dot{x}_{1,sp}\Delta t\right)^{2}}\dot{x}_{2,sp}\left(t\right)+\\
 &  & \frac{1}{h_{2}\left(\bfx_{sp}\left(t\right)+\bfe_{1}\dot{x}_{1,sp}\Delta t\right)^{2}}\int_{0}^{\Delta t}\rmd t'\tilde{F}_{x_{2}}\left(\bfx_{sp}+\bfe_{1}\dot{x}_{1,sp}t',\dot{x}_{1,sp}\right)~,\\
\dot{x}_{3,sp}\left(t+\Delta t\right) & = & \frac{h_{3}\left(\bfx_{sp}\left(t\right)\right)^{2}}{h_{3}\left(\bfx_{sp}\left(t\right)+\bfe_{1}\dot{x}_{1,sp}\Delta t\right)^{2}}\dot{x}_{3,sp}\left(t\right)+\\
 &  & \frac{1}{h_{3}\left(\bfx_{sp}+\bfe_{1}\dot{x}_{1,sp}\Delta t\right)^{2}}\int_{0}^{\Delta t}\rmd t'\tilde{F}_{x_{3}}\left(\bfx_{sp}+\bfe_{1}\dot{x}_{1,sp}t',\dot{x}_{1,sp}\right)~.
\end{array}\right.
\end{eqnarray}
Similarly, analytical solutions $\Theta_{2}$ ($\Theta_{3}$) for
$H_{2}$ ($H_{3}$) can be also derived when $\partial h_{2}\left(\bfx\right)/\partial x_{2}=\partial h_{3}\left(\bfx\right)/\partial x_{3}=0$.
Finally, we can compose these analytical solutions to obtain explicit
symplectic integration algorithms for the entire system. For example,
a first order scheme can be constructed as 
\begin{eqnarray}
\Theta_{1}\left(\Delta t\right)=\Theta_{E}\left(\Delta t\right)\Theta_{B}\left(\Delta t\right)\Theta_{x}\left(\Delta t\right)\Theta_{y}\left(\Delta t\right)\Theta_{z}\left(\Delta t\right)~,
\end{eqnarray}
and a second order symmetric scheme is 
\begin{eqnarray}
\Theta_{2}\left(\Delta t\right) & = & \Theta_{x}\left(\Delta t/2\right)\Theta_{y}\left(\Delta t/2\right)\Theta_{z}\left(\Delta t/2\right)\Theta_{B}\left(\Delta t/2\right)\Theta_{E}\left(\Delta t\right)\nonumber \\
 &  & \Theta_{B}\left(\Delta t/2\right)\Theta_{z}\left(\Delta t/2\right)\Theta_{y}\left(\Delta t/2\right)\Theta_{x}\left(\Delta t/2\right)~.
\end{eqnarray}
An algorithm with order $2(l+1)$ can be constructed in the following
way, 
\begin{eqnarray}
\Theta_{2(l+1)}(\Delta t) & = & \Theta_{2l}(\alpha_{l}\Delta t)\Theta_{2l}(\beta_{l}\Delta t)\Theta_{2l}(\alpha_{l}\Delta t)~,\\
\alpha_{l} & = & 1/(2-2^{1/(2l+1)})~,\\
\beta_{l} & = & 1-2\alpha_{l}~.
\end{eqnarray}

\subsection{High-order explicit structure-preserving geometric PIC algorithm
in a cylindrical mesh \label{Sec2cylindrical}}

Magnetic fusion plasmas are often confined in the toroidal geometry,
for which the cylindrical coordinate system is convenient. We now
present the high-order explicit structure-preserving geometric PIC
algorithm in a cylindrical mesh. In this coordinate system, the line
element is 
\begin{eqnarray}
\rmd s^{2} & = & \left(\rmd r\right)^{2}+\left(\frac{r+R_{0}}{R_{0}}\rmd(R_{0}\xi)\right)^{2}+\left(\rmd z\right)^{2}~,
\end{eqnarray}
where $R_{0}$ is a fixed radial length, $r+R_{0}$ is the radius
in the standard cylindrical coordinate system, and $R_{0}\xi$ is
the polar angle coordinate normalized by $1/R_{0}$. For typical applications
in tokamak physics, $R_{0}$ is the major radius and $\xi$ is called
toroidal angle. To simplify the notation, we also refer to $(r,R_{0}\xi,z)$
as $(x,y,z)$.

For non-relativistic particles in the cylindrical mesh, if the discrete
velocity in \EQ{EqnDV} is changed to 
\begin{eqnarray}
\bfv_{sp,l} & = & \frac{\bfx_{sp,l+1}-\bfx_{sp,l}}{\Delta t}\bfha\left(\bfx_{sp,l+1}\right)~,
\end{eqnarray}
then the 1st-order scheme given by \EQ{EqnDLF} will be explicit.
To construct an explicit 2nd-order scheme, the 2nd-order action integral
can be chosen as 
\begin{eqnarray}
 &  & \calA_{d2}=\sum_{s,p,l}\frac{1}{2}\Big(L_{sp}\left(m_{s},\bfv_{sp,2l}\right)+L_{sp}\left(m_{s},\bfv_{sp,2l+1}^{*}\right)+\nonumber \\
 &  & q_{s}\left(2\overline{\bfv_{sp}\cdot\bfA}\left(\bfx_{sp,2l},\bfx_{sp,2l+1},l\right)+2\overline{\bfv_{sp}^{*}\cdot\bfA}\left(\bfx_{sp,2l+1},\bfx_{sp,2l+2},l\right)-2\phi\left(\bfx_{sp,2l+2},l+1\right)\right)\Big)\nonumber \\
 &  & +\frac{1}{2}\sum_{J}\left(\left(\left(-\frac{\bfA_{J,l+1}-\bfA_{J,l}}{\Delta t}-\left(\nabla_{d}\phi\right)_{J,l+1}\right)\frac{1}{\bfha\left(\bfx_{J}\right)}\right)^{2}-\left(\frac{\left(\CURLD\bfA\right)_{K,l}}{\bfhb\left(\bfx_{K}\right)}\right)^{2}\right)\bfhc\left(\bfx_{J}\right)~,
\end{eqnarray}
where 
\begin{eqnarray*}
 &  & \bfv_{sp,2l}=2\frac{\bfx_{2l+1}-\bfx_{2l}}{\Delta t}\bfha\left(\bfx_{sp,2l+1}\right)~,\\
 &  & \bfv_{sp,2l+1}^{*}=2\frac{\bfx_{2l+2}-\bfx_{2l+1}}{\Delta t}\bfha\left(\bfx_{sp,2l+1}\right)~,\\
 &  & \overline{\bfv_{sp}^{*}\cdot\bfA}\left(\bfx_{sp,2l+1},\bfx_{sp,2l+2},l\right)=\int_{0}^{1}\rmd\tau\\
 &  & \left[\frac{x_{1,sp,2l+2}-x_{1,sp,2l+1}}{\Delta t},\frac{x_{2,sp,2l+2}-x_{2,sp,2l+1}}{\Delta t},\frac{x_{3,sp,2l+2}-x_{3,sp,2l+1}}{\Delta t}\right]\\
 &  & \cdot\left[\begin{array}{l}
A_{x_{1},l}\left(x_{1,sp,2l+1}+\tau\left(x_{1,sp,2l+2}-x_{1,sp,2l+1}\right),x_{2,sp,2l+2},x_{3,sp,2l+2}\right),\\
A_{x_{2},l}\left(x_{1,sp,2l+1},x_{2,sp,2l+1}+\tau\left(x_{2,sp,2l+2}-x_{2,sp,2l+1}\right),x_{3,sp,2l+2}\right),\\
A_{x_{3},l}\left(x_{1,sp,2l+1},x_{2,sp,2l+1},x_{3,sp,2l+1}+\tau\left(x_{3,sp,2l+2}-x_{3,sp,2l+1}\right)\right)
\end{array}\right]~.
\end{eqnarray*}
Taking the following discrete variation yields discrete time-advance
rules, 
\begin{eqnarray}
\frac{\partial\calA_{d2}}{\partial\bfx_{2l}} & = & 0~,\\
\frac{\partial\calA_{d2}}{\partial\bfx_{2l+1}} & = & 0~,\\
\frac{\partial\calA_{d2}}{\partial\bfA_{J,l}} & = & 0~,\\
\frac{\partial\calA_{d2}}{\partial\phi_{J,l}} & = & 0~.
\end{eqnarray}
The explicit expressions for these time-advance are listed in Appendix
\ref{SecESPIC}. Higher-order discrete action integral for building
explicit schemes can be also derived using a similar technique, or
using the splitting method described in Sec.\,\ref{Sec2XQ-bracket}.

In addition, we have also implemented the 1st-order relativistic charge-conserving
geometric PIC in the cylindrical mesh. However the particle pusher
in the relativistic scheme is implicit, and it is about 4 times slower
than the explicit scheme. When studying relativistic effects, the
implicit relativistic algorithm can be applied without significantly
increasing the computational cost.

\section{Whole-device 6D kinetic simulations of tokamak physics}

\label{Sec3} The high-order explicit structure-preserving geometric
PIC algorithm in a cylindrical mesh described in Sec.\,\ref{Sec2cylindrical}
has been implemented in the \textsl{SymPIC} code, designed for high-efficiency
massively-parallel PIC simulations in modern clusters. The \textsl{OpenMP-MPI}
version of \textsl{SymPIC} is available at \url{https://github.com/JianyuanXiao/SymPIC/}.
The algorithm and code have been used to carry out the first-ever
whole-device 6D kinetic simulations of tokamak physics. In this section,
we report simulation results using machine parameters similar to those
of the Alcator C-Mod tokamak \citep{hutchinson1994first,greenwald1997h}.
Two physics problems are studied, the self-consistent kinetic steady
state and kinetic ballooning mode instabilities in the self-consistent
kinetic steady state.

\subsection{Axisymmetric self-consistent kinetic steady state in a tokamak \label{SecTK2D}}

\begin{figure}
\begin{centering}
\subfloat[Locations of PF coils and poloidal field lines.]{\includegraphics[width=0.43\linewidth]{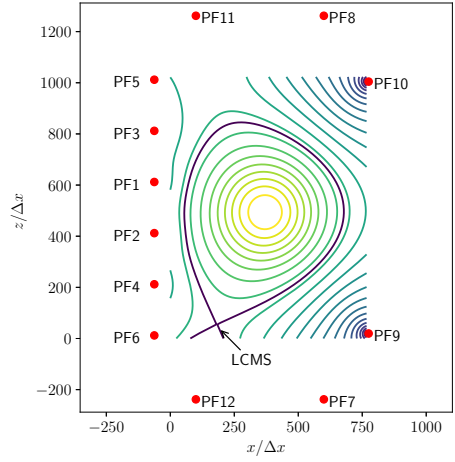}

}
\par\end{centering}
\begin{centering}
\subfloat[Initial plasma density profile ($5\EXP{19}\mathrm{m}^{-3})$.]{\includegraphics[width=0.56\linewidth]{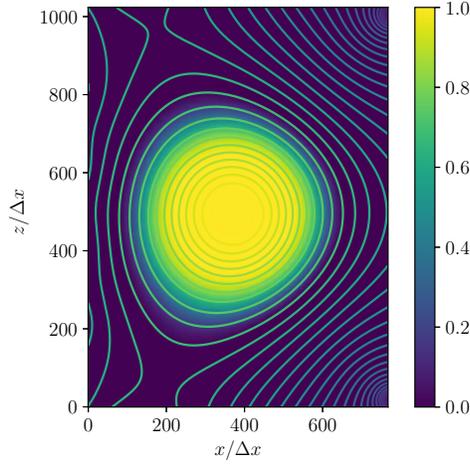}

}
\par\end{centering}
\caption{Poloidal field and initial plasma density profile.}
\label{FigTKMK} 
\end{figure}

Kinetic equilibrium is the starting point for analytical and numerical
studies of kinetic instabilities and associated transport phenomena.
Because no self-consistent kinetic equilibrium is known in the tokamak
geometry, most previous studies adopted non-self-consistent distributions
as assumed kinetic equilibria, especially for simulations based on
the $\delta f$-method. Here, we numerically obtain an axisymmetric
self-consistent kinetic steady state for a small tokamak using parameters
similar to those of the Alcator C-Mod tokamak \citep{hutchinson1994first,greenwald1997h}.
The machine parameters are tabulated in Table \ref{TabTKMK}. The
device is numerically constructed with Poloidal Field (PF) coils displayed
in Fig.\,\ref{FigTKMK}(a).

\begin{table}
\centering %
\begin{tabular}{|c|c|c|c|c|}
\hline 
Major radius $R_{0}$  & Minor radius $a$  & Plasma current $I_{p}$  & Edge safe factor $q_{95}$  & %
\begin{minipage}[c][1.5cm]{3.5cm}%
Toroidal magnetic field at $R_{0}$ ($B_{0}$)%
\end{minipage}\tabularnewline
\hline 
0.69m  & 0.21m  & 0.54MA  & About 3.5  & 4.2T \tabularnewline
\hline 
\end{tabular}\caption{Machine parameters of the simulated tokamak. This set of parameters
is similar to those of the Alcator C-Mod tokamak.}
\label{TabTKMK} 
\end{table}

\begin{figure}
\begin{centering}
\includegraphics[width=0.9\linewidth]{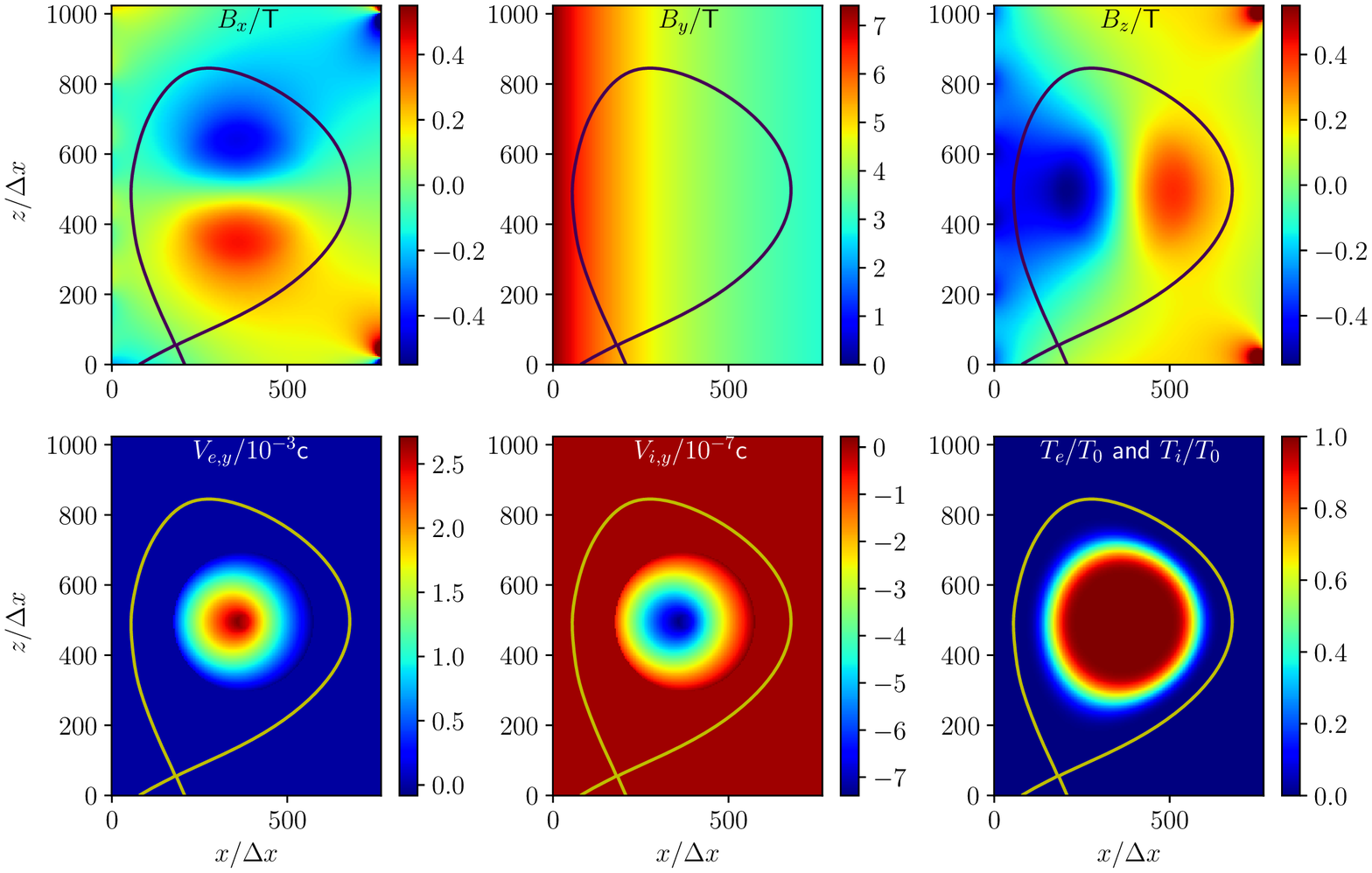} 
\par\end{centering}
\caption{Tokamak external magnetic field and initial profiles of flow velocities
in the $\bfe_{y}$ direction and temperatures.}
\label{FigTKMKV} 
\end{figure}

\begin{figure}
\begin{centering}
\includegraphics[width=0.9\linewidth]{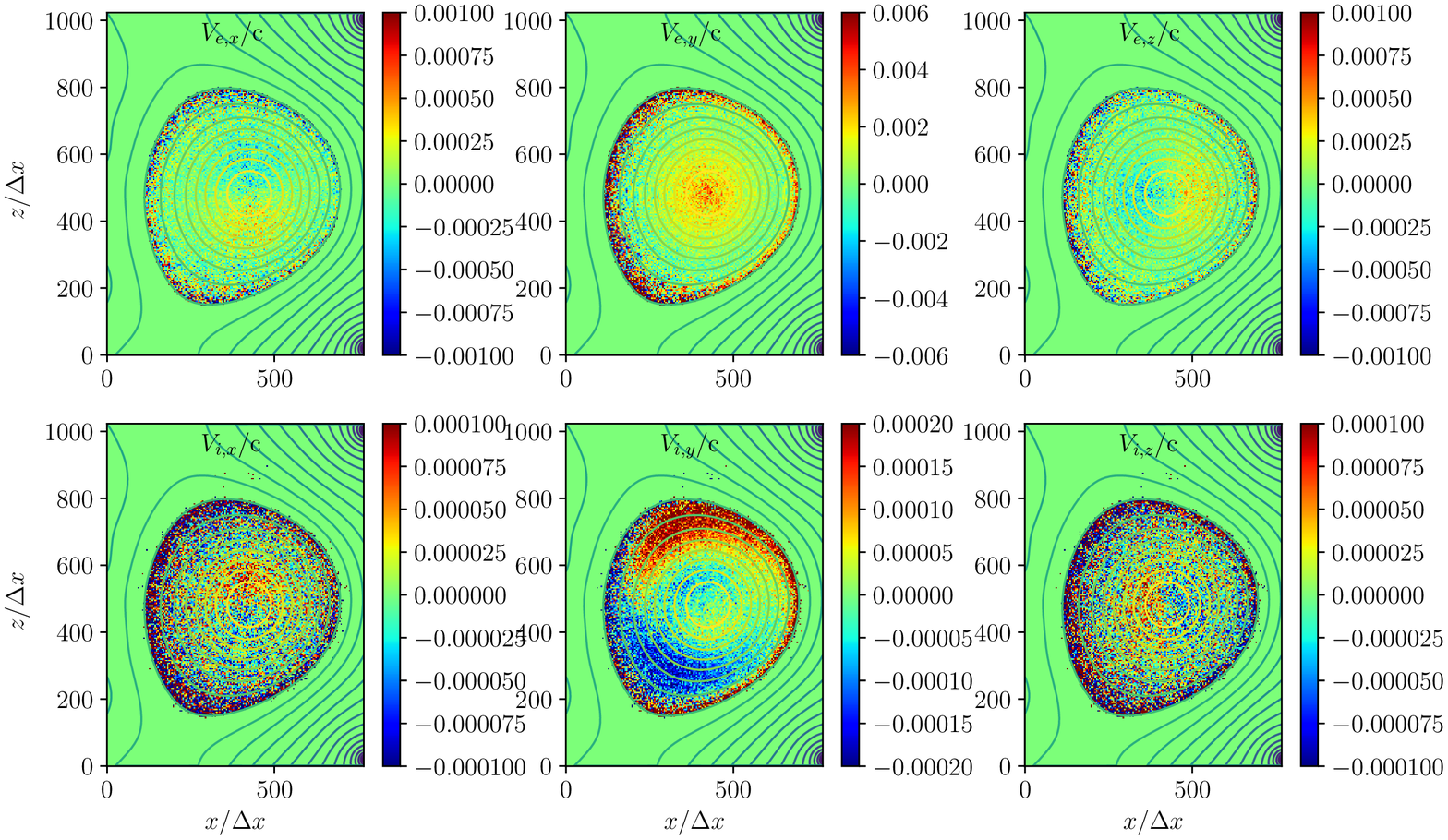} 
\par\end{centering}
\caption{Flow velocity distributions for electrons and ions at the steady state.}
\label{FigVEN} 
\end{figure}

\begin{figure}
\begin{centering}
\includegraphics[width=0.654545\linewidth]{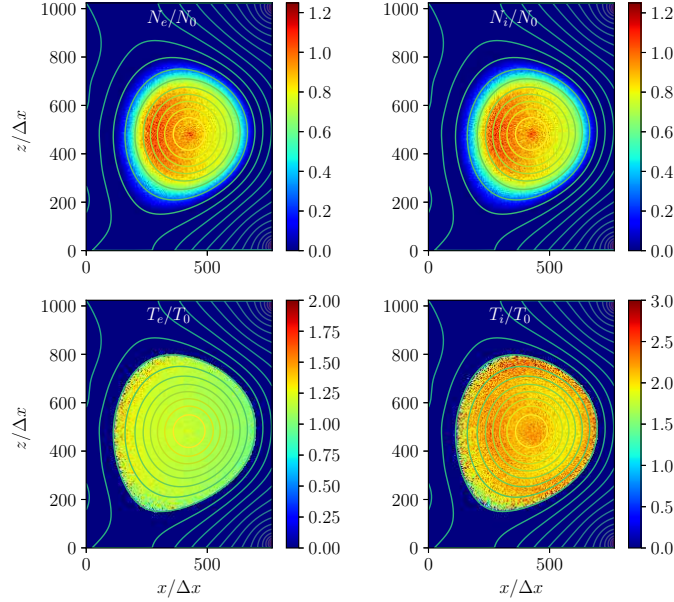} 
\par\end{centering}
\caption{Density and temperature profiles of electrons and ions at the steady
state.}
\label{FigVIN} 
\end{figure}

\begin{figure}
\begin{centering}
\includegraphics[width=0.9\linewidth]{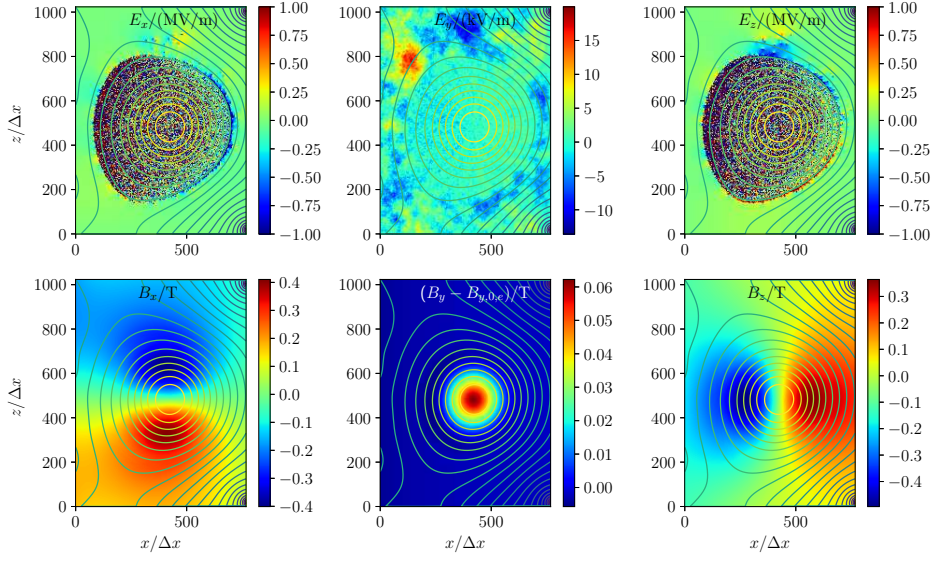} 
\par\end{centering}
\caption{Electromagnetic field profiles at the steady state.}
\label{FigEB} 
\end{figure}

\begin{figure}
\begin{centering}
\includegraphics[width=0.9\linewidth]{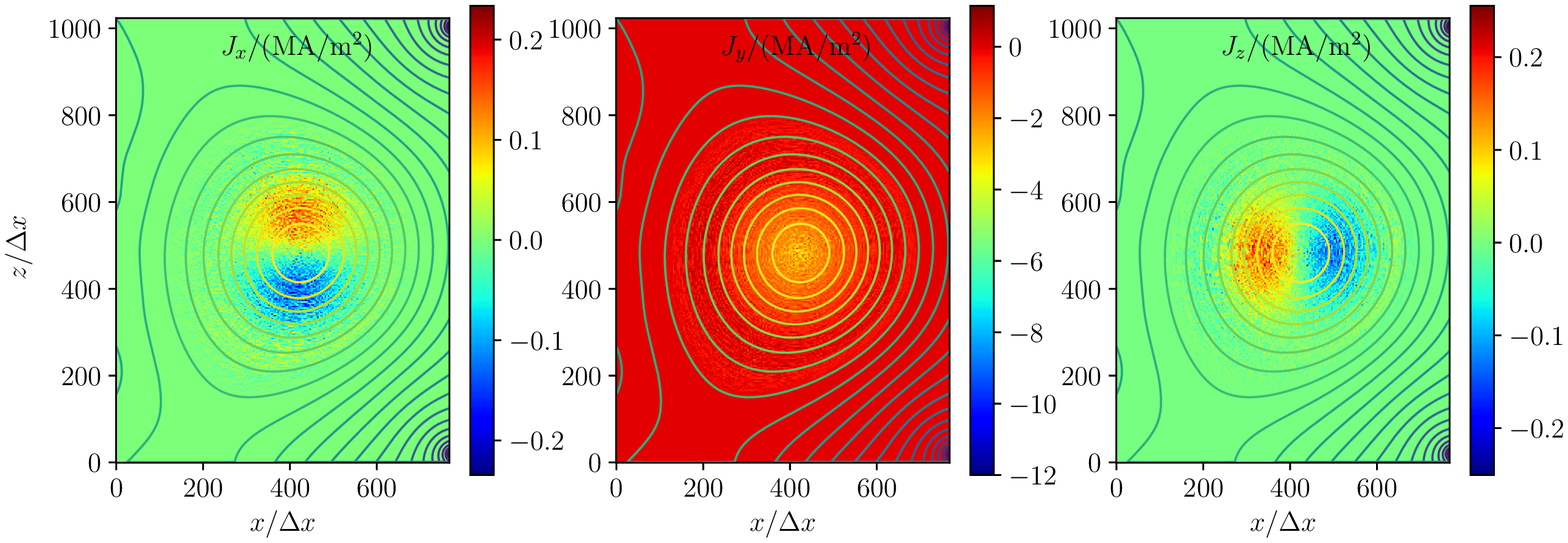} 
\par\end{centering}
\caption{Current profiles at the steady state.}
\label{FigJ} 
\end{figure}

At the beginning of the simulation, non-equilibrium distributions
for deuterium ions and electrons are loaded into the device. The initial
density profile in the poloidal plane is shown in Fig.\,\ref{FigTKMK}(b).
The 2D profiles of the external field and initial velocity and temperature
profiles for both species are plotted in Fig.\,\ref{FigTKMKV}. Simulation
parameters are chosen as 
\begin{eqnarray*}
\Delta x & = & 8\EXP{-4}\rmm,\,\,\Delta t=0.1\Delta x/\rmc,\,\,R_{0}=500\Delta x~,\\
T_{e,0} & = & T_{i,0}=T_{0}=800\textrm{eV},\,\,n_{e,0}=n_{i,0}=5.0\EXP{19}\mathrm{m}^{-3},\\
m_{i} & = & 3672m_{e}=3.34\EXP{-27}\mathrm{kg},\\
J_{PF,i} & = & 67.9\mathrm{kA},\quad\textrm{for }1\leq i\leq6~,\\
J_{PF,7} & = & J_{PF,8}=211.1\mathrm{kA},\,\,J_{PF,9}=J_{PF,10}=238.1\mathrm{kA}~,\\
J_{PF,11} & = & -189.9\mathrm{kA},\,\,J_{PF,12}=-211.1\mathrm{kA}~.
\end{eqnarray*}
Detailed calculation of the external magnetic field, initial particle
distributions, and boundary setup are outlined in \APP{SecAPP}.
Simulations show that the system reaches a steady state after $1.6\EXP7$
time-steps. At this time, the amplitude of magnetic perturbation at
the middle plane of the edge ($x=0.48$m and $z=0.41$m) is smaller
than $B_{0}/200$, and the oscillation of position of plasma core
is very small. Therefore, we can treat this state as a steady state.
One such calculation requires about $1.3\EXP{5}$ core-hours on the
Tianhe 3 prototype cluster. Profiles of flow velocities, densities,
temperatures of electrons and ions as well as electromagnetic field
at the steady state are shown in Figs.~\ref{FigVEN}-\ref{FigJ}.



\begin{figure}
\includegraphics[width=0.49\linewidth]{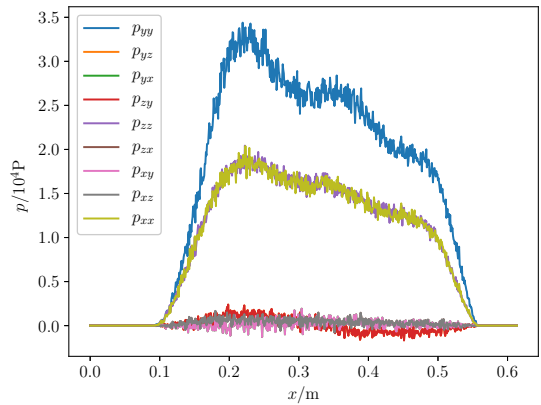}\caption{Pressure tensor profile at $z=0.41\mathrm{m}$.}
\label{FigPresT} 
\end{figure}

\begin{figure}
\begin{centering}
\subfloat[$\bfe_{x}$ direction]{\begin{centering}
\includegraphics[width=0.49\linewidth]{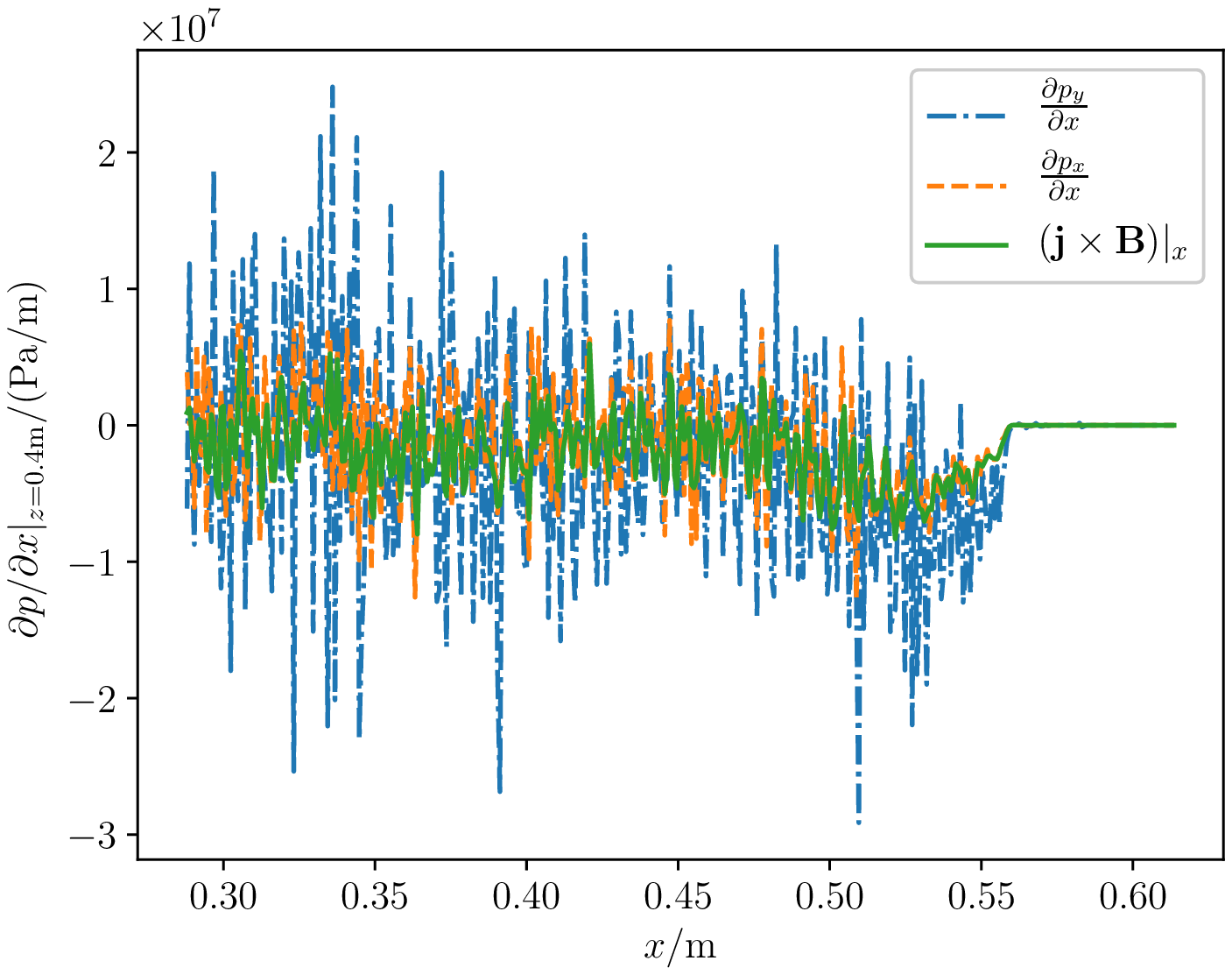} 
\par\end{centering}
}
\par\end{centering}
\begin{centering}
\subfloat[$\bfe_{z}$ direction]{\begin{centering}
\includegraphics[width=0.49\linewidth]{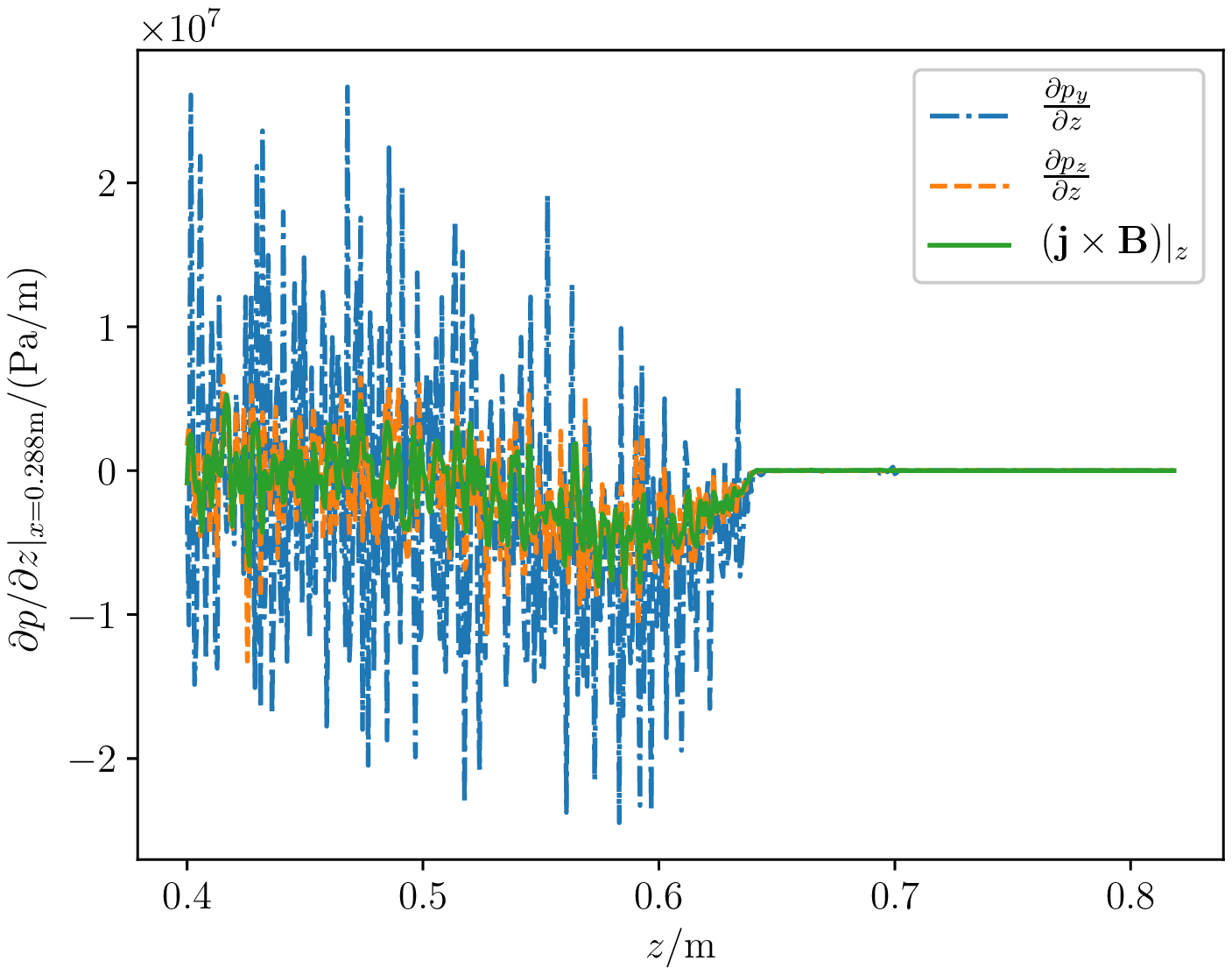} 
\par\end{centering}
}
\par\end{centering}
\caption{Pressure gradient $\nabla\cdot\boldsymbol{p}$ and Lorentz force $\bfJ\times\bfB$
in the $\bfe_{x}$ (a) and $\bfe_{z}$ (b) directions.}
\label{FigPresG} 
\end{figure}

From Fig.\,\ref{FigVEN}, the flow velocity of ions at the steady
state is in the range of $10$km/s, which is consistent with experimental
observations \citep{Ince-Cushman2009,Rice2009} and theoretical calculations
\citep{Guan2013,Guan2013a}. This ion flow is much slower than the
thermal velocity of ions and thus is negligible in the force balance
for the steady state, which can be written as

\begin{eqnarray}
\bfJ\times\bfB-\nabla\cdot\boldsymbol{p}=0~.\label{eq:balance}
\end{eqnarray}
Here, $\boldsymbol{p}$ is the pressure tensor. Equation (\ref{eq:balance})
is obtained by the familiar procedure of taking the second moment
of the Vlasov equation, subtracting the flow velocity and summing
over species. From the simulation data, the $p_{ij}$ component of
the pressure tensor is calculated as

\begin{align}
p_{ij,J} & =\sum_{\bfx_{sp}\in\textrm{grid }J}m_{s}v'_{i,sp}v'_{j,sp}~,\quad\textrm{for }i,j\textrm{ in }\left\{ x,y,z\right\} ~,\\
\bfv'_{sp} & \equiv\bfv_{sp}-\left(\sum_{\bfx_{sp}\in\textrm{grid }J}\bfv_{sp}\right)/\left(\sum_{\bfx_{sp}\in\textrm{grid }J}1\right)~.
\end{align}
The profile of pressure tensor at the steady state at $z=0.41$m is
shown in \FIG{FigPresT}. Clearly, the pressure tensor is predominately
diagonal and anisotropic with $p_{yy}>p_{xx}\approx p_{zz}$. Note
that the pressure is almost isotropic in the poloidal plane, which
indicates that it is valid to adopt the ideal magnetohydrodynamics
(MHD) model with a scalar pressure for force balance in the 2D tokamak
equilibrium. However, for 3D physics, the effect of pressure anisotropy
needs to be considered. Since the steady state is 2D in space for
the present case, the force balance equation reduces to 
\begin{align}
\frac{\partial p_{xx}}{\partial x} & =\left(\bfJ\times\bfB\right)_{x}\,,\label{eq:pxx}\\
\frac{\partial p_{zz}}{\partial z} & =\left(\bfJ\times\bfB\right)_{z}\,.\label{eq:pzz}
\end{align}
To verify the force balance of the steady state, the four terms in
Eqs.\,(\ref{eq:pxx}) and (\ref{eq:pzz}) are plotted in Fig.\,\ref{FigPresG}.
For comparison, the components of $\partial p_{yy}/\partial x$ and
$\partial p_{yy}/\partial z$ are also plotted. Figure \ref{FigPresG}
shows that the force balance is approximately satisfied for the numerically
obtained steady state, which can be viewed as a self-consistent kinetic
steady state. 

\begin{figure}
\begin{centering}
\includegraphics[width=0.6\linewidth]{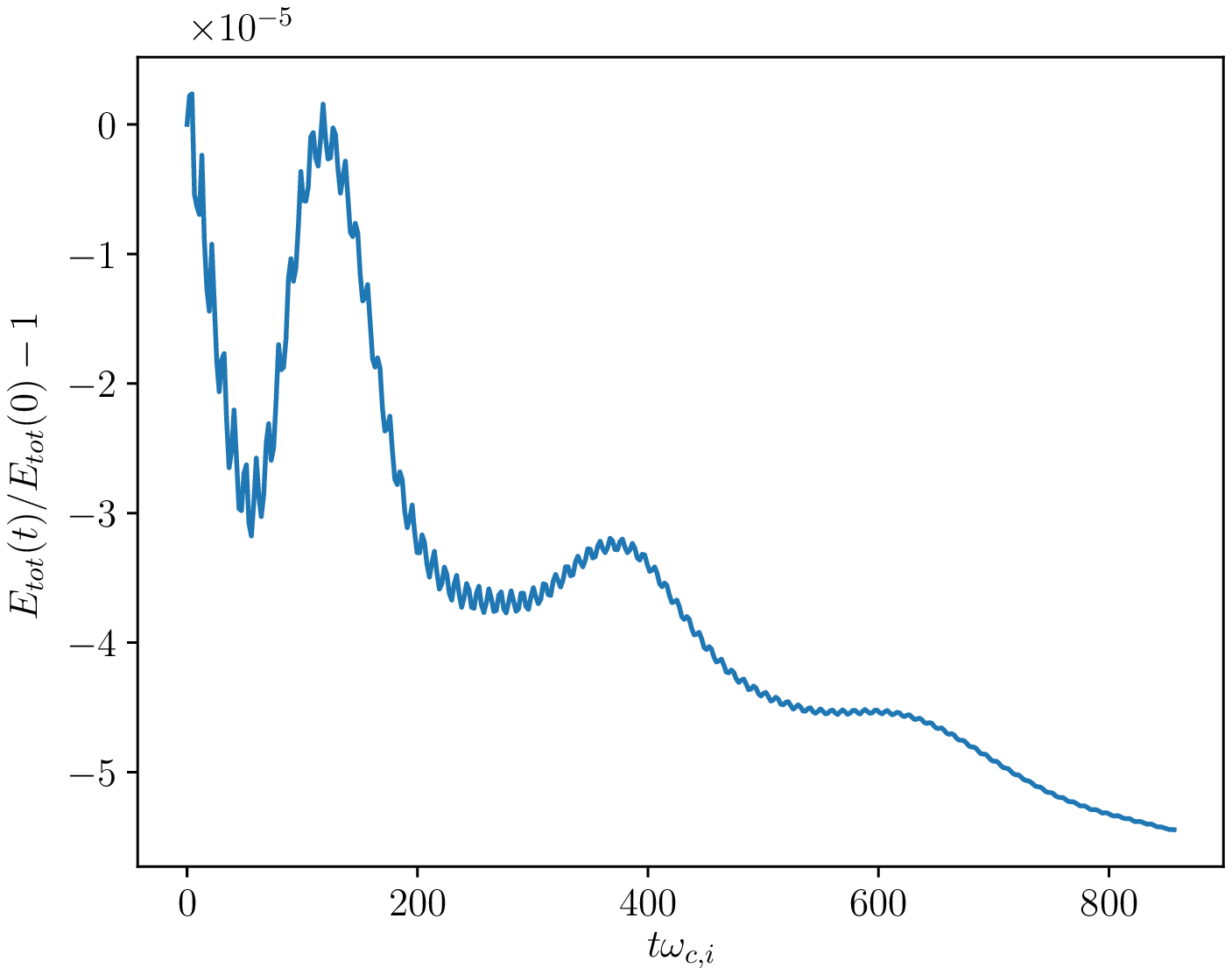} 
\par\end{centering}
\caption{Evolution of total energy.}
\label{FigEne} 
\end{figure}

To verify the energy conservation in the simulation, we have recorded
the time-history of the total energy in \FIG{FigEne}. The total
energy drops a little, because some particles outside the last closed
magnetic surface hit the simulation boundary, and these particles
are removed from the simulation. 


\begin{figure}
\begin{centering}
\includegraphics[width=0.6\linewidth]{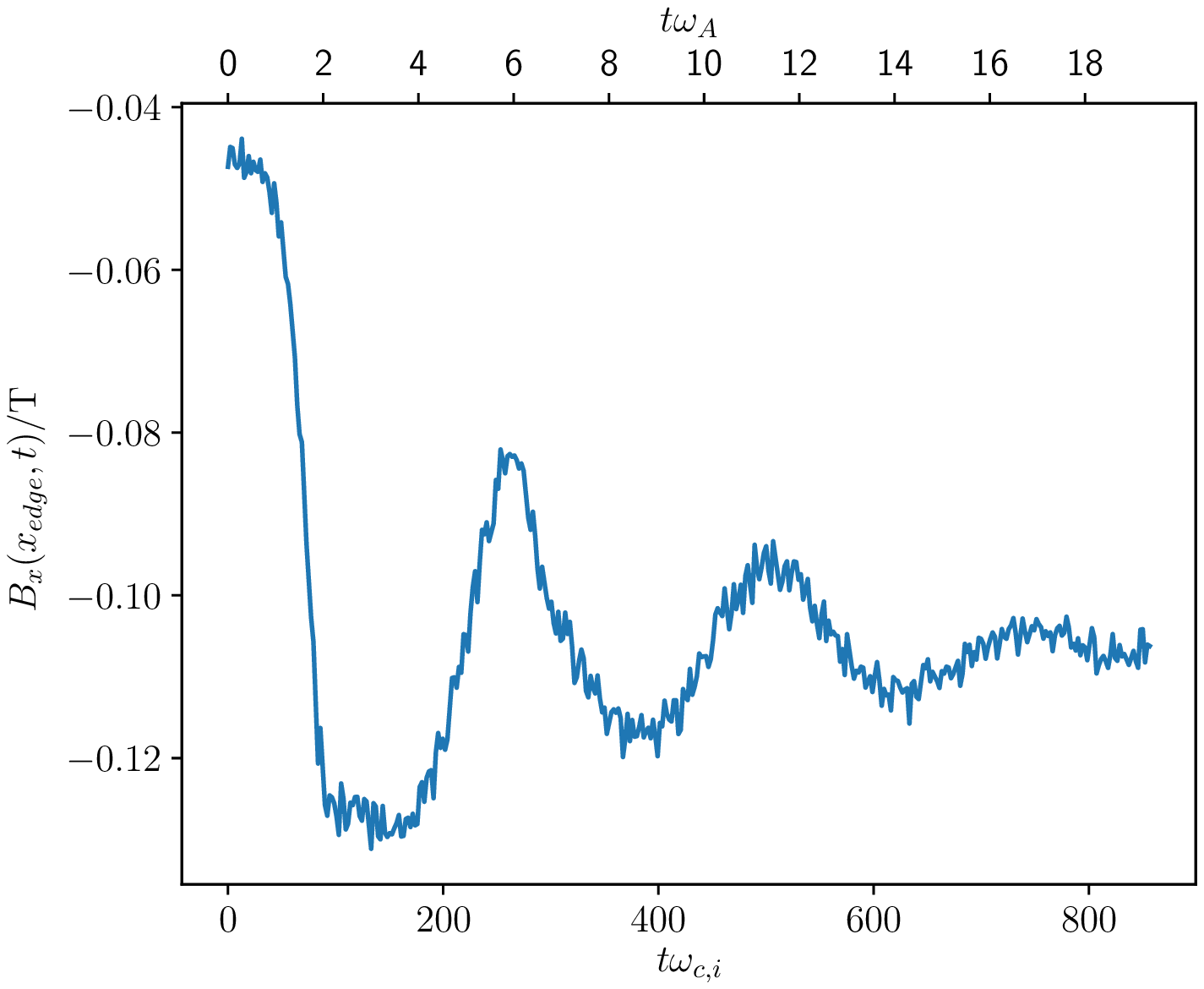} 
\par\end{centering}
\caption{Evolution of $B_{x}(600\Delta x,0,512\Delta x)$.}
\label{FigBext} 
\end{figure}

Before reaching the kinetic steady state, the plasma oscillates in
the poloidal plane. It is expected that this oscillation can be described
as an MHD process whose characteristic velocity is the Alfvén velocity
$v_{A}=B_{0}/\sqrt{\mu_{0}n_{i}m_{i}}$. To observe this oscillation,
we plot in \FIG{FigBext} the evolution of the magnetic field at
$x=600\Delta x$ and $z=512\Delta x$. From the parameters of the
simulation, the characteristic frequency of the oscillation is 
\begin{eqnarray}
\omega_{A}=\frac{v_{A}}{qR_{0}}\sim2.26\EXP{-2}\omega_{c,i}~,
\end{eqnarray}
which agrees with the frequency of the $B_{x}$ oscillation in \FIG{FigBext}.



\subsection{Kinetic ballooning mode in tokamak}

Kinetic Ballooning Mode (KBM) \citep{Tang1980}, characterized by
both electromagnetic perturbations of the MHD type and nontrivial
kinetic effects, plays an important role in tokamak edge physics.
Traditionally, it has been simulated using electromagnetic gyrokinetic
codes such as the Kin-2DEM \citep{Qin98-thesis,Qin1999}, LIGKA \citep{Lauber2007},
GTC \citep{zebin2013gtc,dong2019nonlinear} and GEM \citep{wang2012linear}.
However, for edge plasmas, the gyrokinetic ordering \citep{Hahm88-2670,Brizard1989,Qin1998,Qin99-1575,Qin99-2544,Qin2000,Sugama00-466,Qin2004,QinFields,Qin2007,burby2014hamiltonian,BurbyThesis,Burby2019}
may become invalid under certain parameter regimes for modern tokamaks.
For instance, the characteristic length in the edge of the H-mode
plasma simulated by Wan et al. \citep{wan2012z,wan2013global} can
be as short as about 5 times of the gyroradius of thermal ions, and
in this situation, the gyrokinetic density calculation may be inaccurate.
We have applied \textsl{SymPIC} code developed to carry out the first
whole-device 6D kinetic simulations of the KBM in a tokamak geometry.

\global\long\def\CRIT{\mathrm{crit}}

The machine parameters are the same as in \SEC{SecTK2D}. To trigger
the KBM instability, we increase the plasma density to $n_{0}=1\EXP{20}\mathrm{m}^{-3}$,
and the rest of parameters are 
\begin{eqnarray*}
I_{p} & = & 0.858\mathrm{MA},\,\,T_{e,0}=T_{i,0}=T_{0}=800\textrm{eV}~,\\
\Delta x & = & 3.2\EXP{-3}\rmm,\,\,\Delta t=0.5\Delta x/\rmc_{n},\,\,R_{0}=125\Delta x~,\\
c_{n} & = & r_{c}\mathrm{c},\,\,m_{i}=r_{m}m_{e}=3.34\EXP{-27}\mathrm{kg},\\
J_{PF,i} & = & 144.6\mathrm{kA},\quad\textrm{for }1\leq i\leq6~,\\
J_{PF,7} & = & J_{PF,8}=448.3\mathrm{kA},\,\,J_{PF,9}=J_{PF,10}=506.2\mathrm{kA}~,\\
J_{PF,11} & = & -367.6\mathrm{kA},\,\,J_{PF,12}=-448.3\mathrm{kA}~.
\end{eqnarray*}
Here $r_{m}$ is the mass ratio between the deuterium and electron,
$c_{n}$ is the speed of light in the simulation and $r_{c}$ is the
ratio between $c_{n}$ and the real speed of light in the vacuum $\mathrm{c}$.
For real plasmas, $r_{m}\approx3672$ and $r_{c}=1$. Limited by available
computation power, we reduce $r_{m}$ and $r_{c}$ in some of the
simulations. Such an approximation is valid because the low frequency
ion motion is relatively independent from the mass of electron and
the speed of light, as long as $r_{m}\gg1$ and $\left(\mathrm{c}r_{c}/v_{A}\right)^{2}\gg1.$
In the present work, we take $r_{m}=100$ and $r_{c}=0.16$ to obtain
long-term simulation results. Short-term results for $r_{m}=300$,
$r_{c}=0.5$ and $r_{m}=3672$, $r_{c}=1$ (in this case $\Delta t=0.1\Delta x/\rmc$)
are also obtained for comparison. The simulation domain is a $192\times64\times256$
mesh, where perfect electric conductor is assumed at the boundaries
in the $x$- and $z$-directions, and the periodic boundary is selected
in $y$ direction.

Because of the steep pressure gradient in the edge of the plasma,
the threshold $\beta_{\CRIT}$ for ballooning mode instability is
low. An estimated scaling for $\beta_{\CRIT}$ is \citep{pueschel2008gyrokinetic},
\[
\beta_{\CRIT}=0.6\hat{s}/\left(\frac{2q^{2}R_{0}}{L_{p}}\right)~,
\]
where $\hat{s}$ is the magnetic shear and $L_{p}$ is the pressure
scaling length. For our simulated plasma, $\beta_{\CRIT}\sim1\EXP{-3}$
and $\beta\approx3\EXP{-3}$. We except to observe unstable KBM.

\begin{figure}[htp]
\begin{centering}
\subfloat[Temperatures for electrons $T_{\mathrm{e}}$ and ions $T_{\mathrm{i}}$,
and the safety factor $q$]{\includegraphics[width=0.49\linewidth]{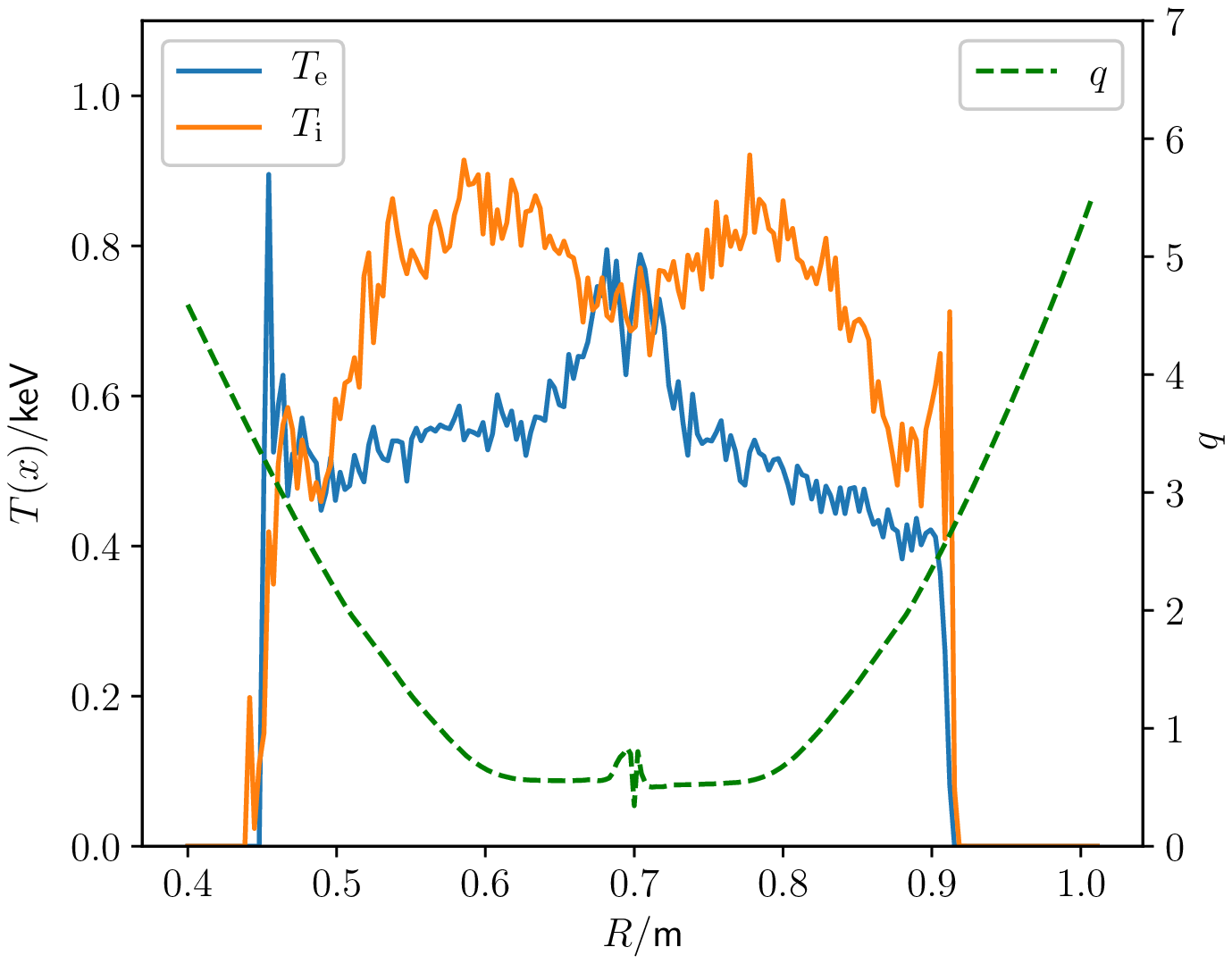} 
}\subfloat[Toroidal current $J_{\mathrm{y}}$, toroidal bulk velocities for electrons
$V_{\mathrm{e,y}}$ and ions $V_{\mathrm{i,y}}$]{\begin{centering}
\includegraphics[width=0.49\linewidth]{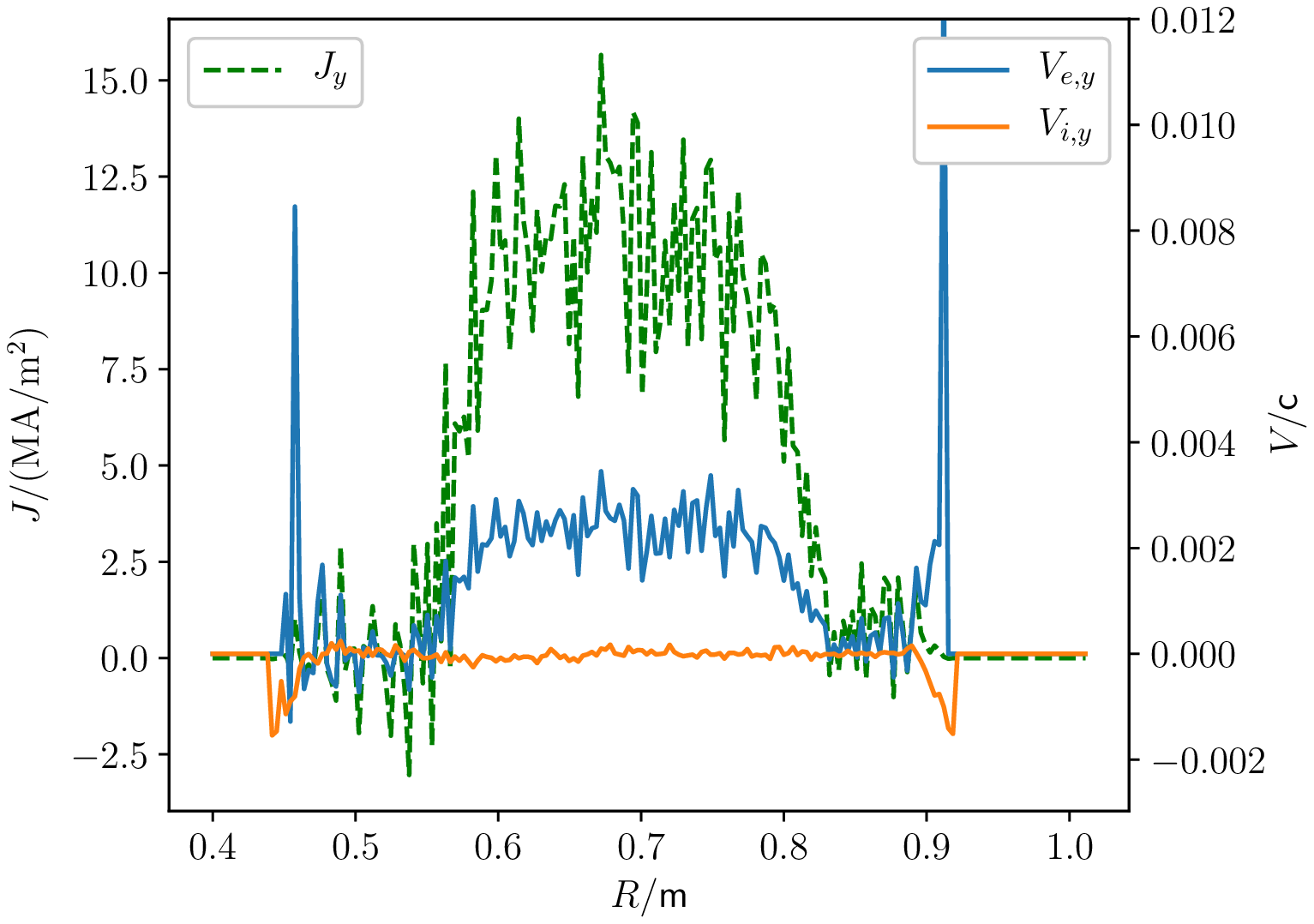} 
\par\end{centering}
}
\par\end{centering}
\begin{centering}
\subfloat[Pressures for electrons $P_{\mathrm{e}}$ and ions $P_{\mathrm{i}}$,
and the number density of electron, here for both electrons and ions
it is found that $P_{i/e,x}\approx P_{i/e,z}$ so only $(P_{i/e,x}+P_{i/e,z})/2$
is plotted.]{\begin{centering}
\includegraphics[width=0.49\linewidth]{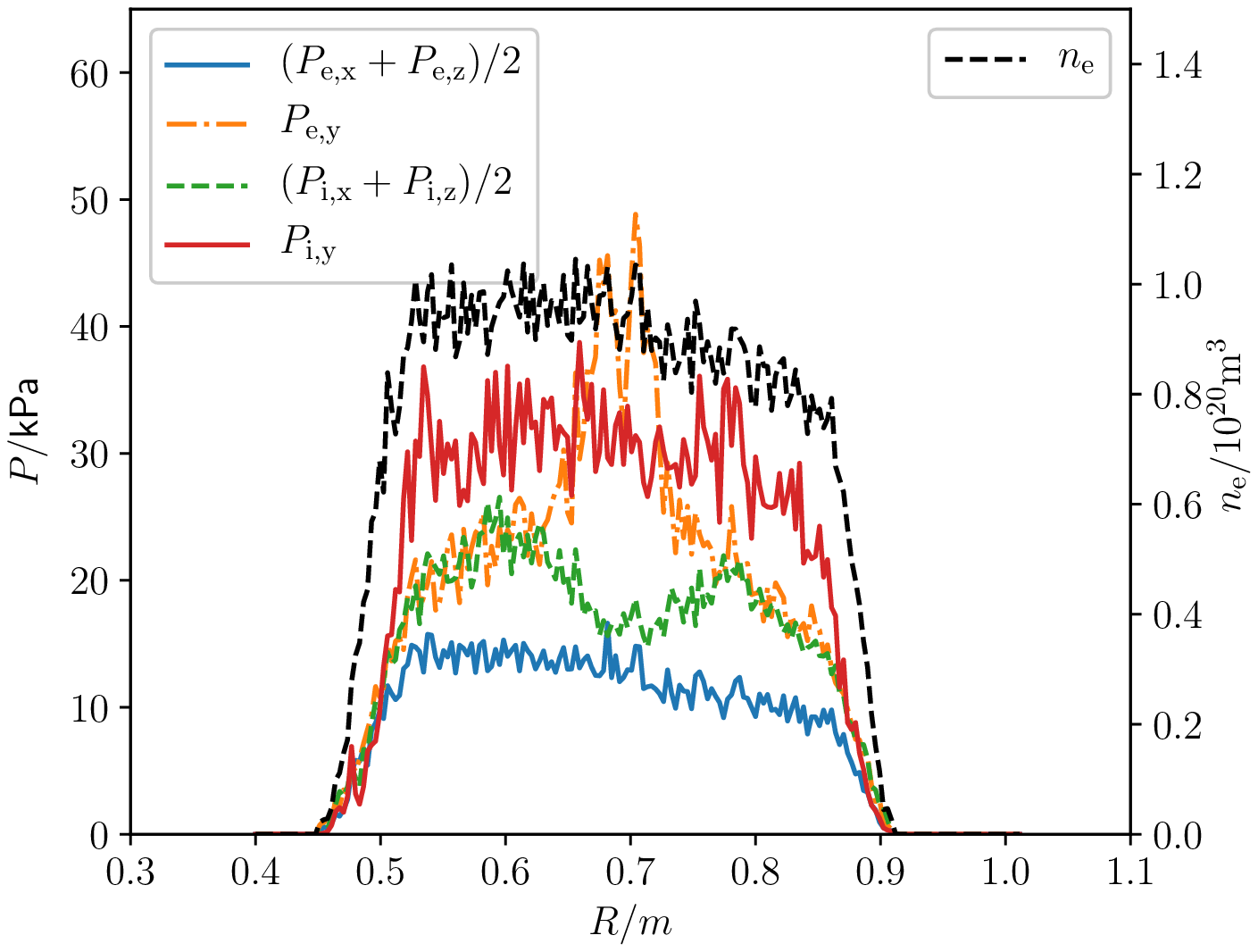} 
\par\end{centering}
}
\par\end{centering}
\caption{Profiles of the kinetic steady state at $z=0.41$m.}
\label{Fig2DST} 
\end{figure}

To obtain a self-consistent kinetic steady state, we first perform
a simulation as described in Sec.\,\ref{SecTK2D} and obtain the
2D kinetic steady state. The profiles of temperature, safety factor,
number density, pressure, toroidal current and bulk velocity of this
steady state at $z=0.41$m are shown in \FIG{Fig2DST}.

\begin{figure}[htp]
\includegraphics[width=0.9\linewidth]{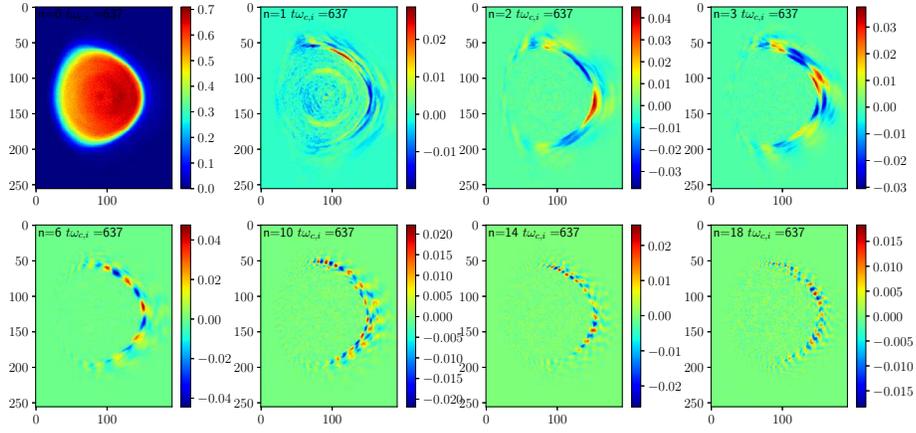}\caption{Mode structures of the electron density at $t=637/\omega_{c,i}=2.85a/v_{t,i}$.}
\label{FigMode16} 
\end{figure}

\begin{figure}[htp]
\includegraphics[width=0.9\linewidth]{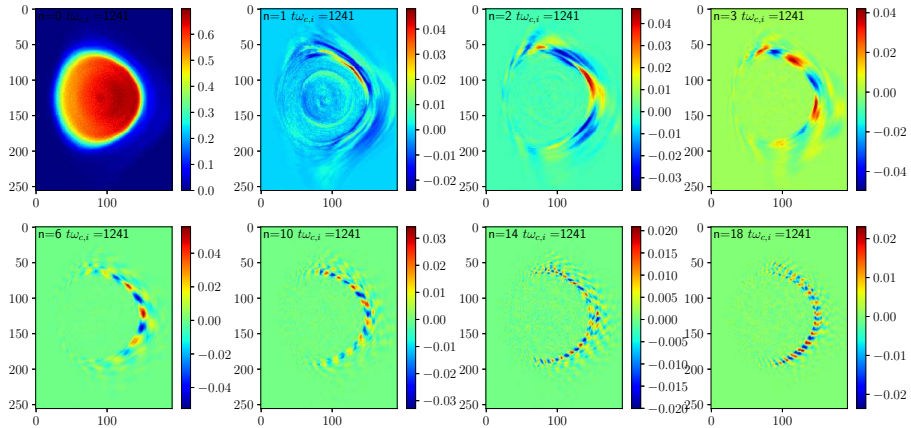}\caption{Mode structures of the electron density at $t=1241/\omega_{c,i}=5.55a/v_{t,i}$.}
\label{FigMode17} 
\end{figure}

\begin{figure}[htp]
\begin{centering}
\includegraphics[width=0.9\linewidth]{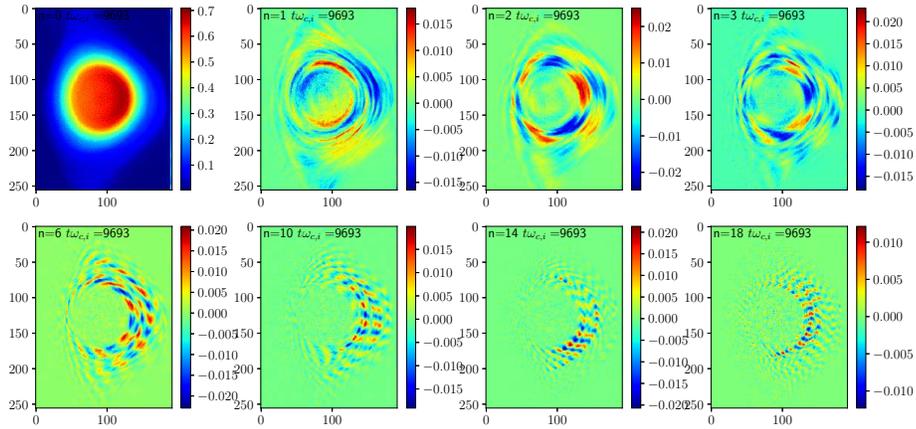} 
\par\end{centering}
\caption{Mode structure of the electron density at $t=9693/\omega_{c,i}=43.35a/v_{t,i}$.}
\label{FigMode18} 
\end{figure}

\begin{figure}[htp]
\begin{centering}
\includegraphics[width=0.6\linewidth]{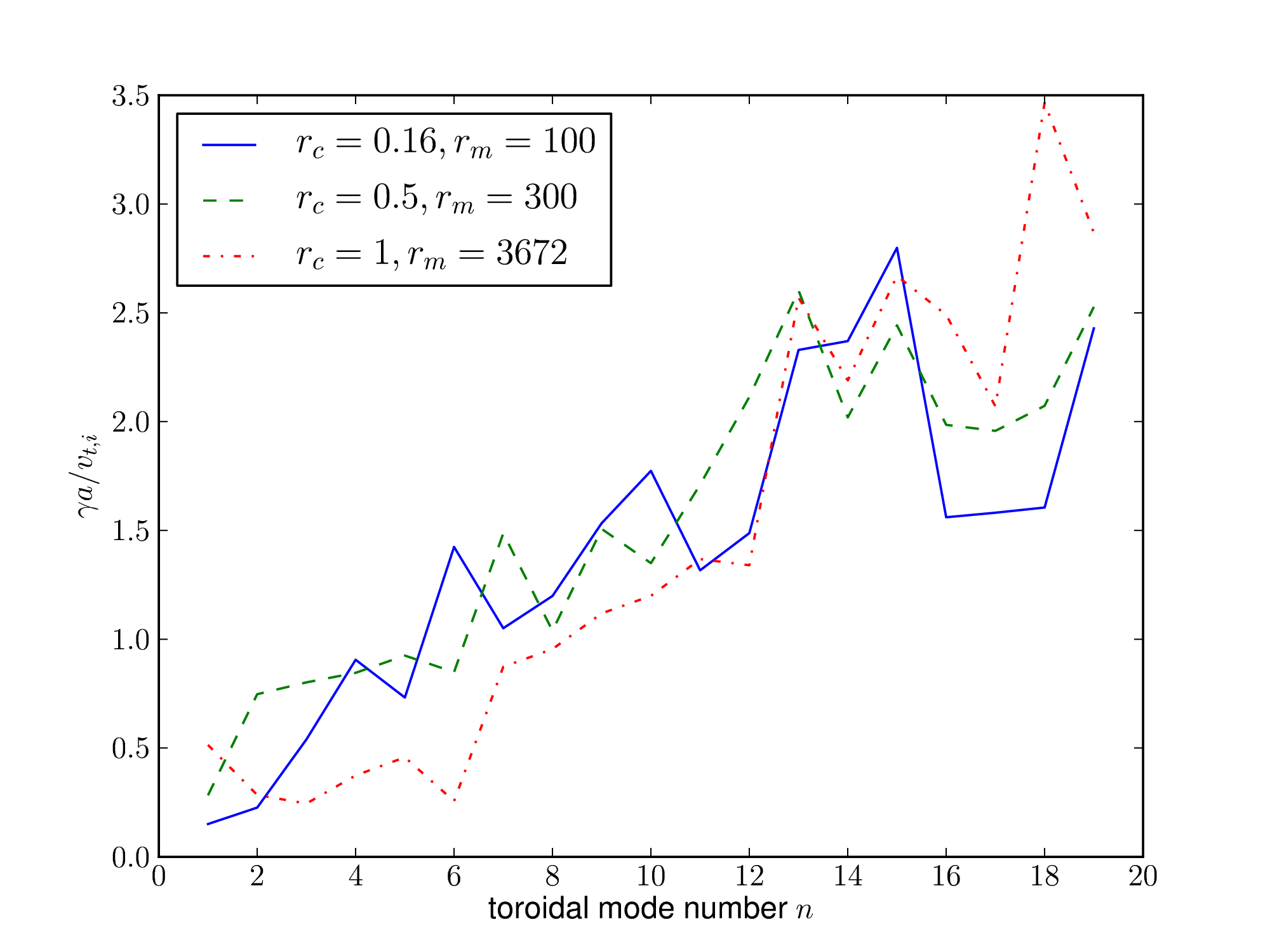} 
\par\end{centering}
\caption{Growth rate as a function of toroidal mode number for different values
of $r_{c}$ and $r_{m}$.}
\label{FigModeGAM} 
\end{figure}

A random perturbation is then added in as the initial condition of
the 6D simulation. The total simulation time is $t_{a}=1.5\EXP{6}\Delta t\approx10000\omega_{c,i}^{-1}\approx5\EXP{-5}$s.
For one such simulation it takes about $3\EXP{5}$ core-hours on the
Tianhe 3 prototype cluster. The resulting mode structures of ion density
for toroidal mode number $n=1,2,3,6,10,14,18$ are shown in Figs.
\ref{FigMode16}, \ref{FigMode17} and \ref{FigMode18} for $t\omega_{c,i}=637,\,\,1241$
and $9693.$ It is clear that the unstable modes are triggered at
the edge of the plasma, and the ballooning structure can be observed
for modes with large $n$. The growth rate as a function of $n$ is
plotted in \FIG{FigModeGAM}. For comparison, the grow rate obtained
using $r_{c}=0.5$, $r_{m}=300$ and $r_{c}=1$, $r_{m}=3672$ are
also plotted. It is clear that the growth rate has little correlation
with the reduction of $r_{c}$ and $r_{m}$. Figure \ref{FigModeGAM}
shows that the growth rate increases with $n,$ consistent with the
early gyrokinetic simulation results obtained using the Kin-2DEM eigenvalue
code \citep{Qin98-thesis,Qin1999}. Because the number of grids in
the toroidal direction is 64 and the width of interpolating function
is 4 times the grid size, the results for modes with $n>16$ may not
be accurate. The results displayed here are thus preliminary. In the
next step, we plan to perform a larger scale simulation with more
realistic 2D equilibria to obtain improved accuracy.

\begin{figure}[htp]
\begin{centering}
\subfloat[Short-term evolution.]{\begin{centering}
\includegraphics[width=0.49\linewidth]{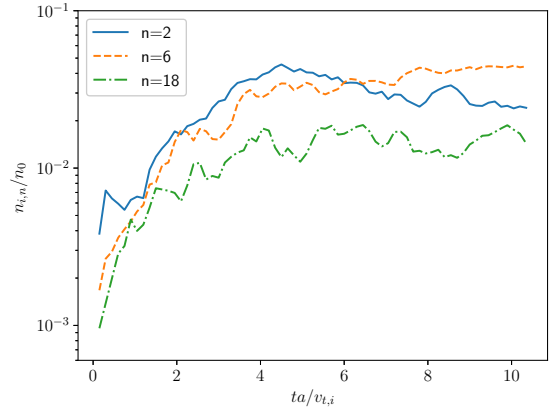} 
\par\end{centering}
}
\par\end{centering}
\begin{centering}
\subfloat[Long-term evolution.]{\includegraphics[width=0.49\linewidth]{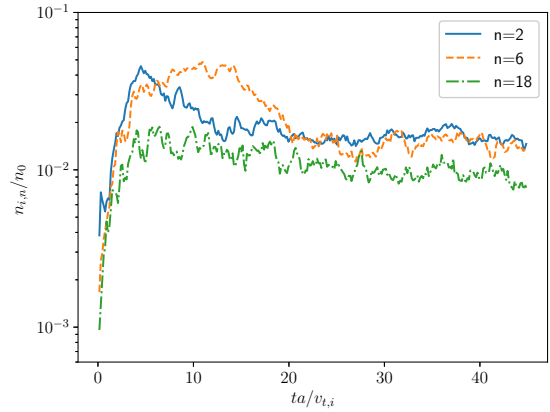}

}
\par\end{centering}
\caption{Time-history of ion densities for different toroidal mode number $n$.}
\label{FigModeAMP} 
\end{figure}

\begin{figure}[htp]
\begin{centering}
\subfloat[]{\begin{centering}
\includegraphics[width=0.49\linewidth]{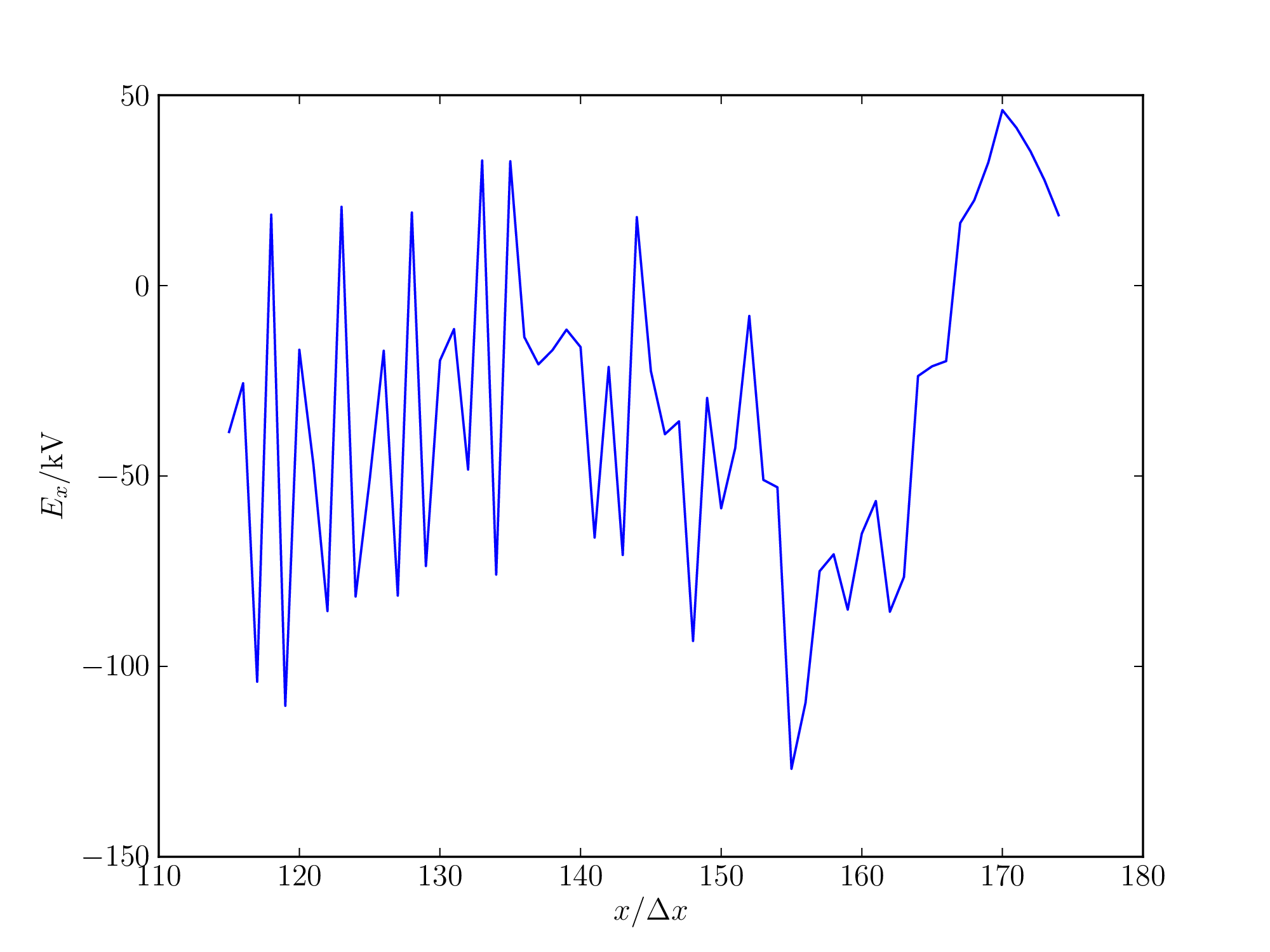} 
\par\end{centering}
}
\par\end{centering}
\begin{centering}
\subfloat[]{\begin{centering}
\includegraphics[width=0.49\linewidth]{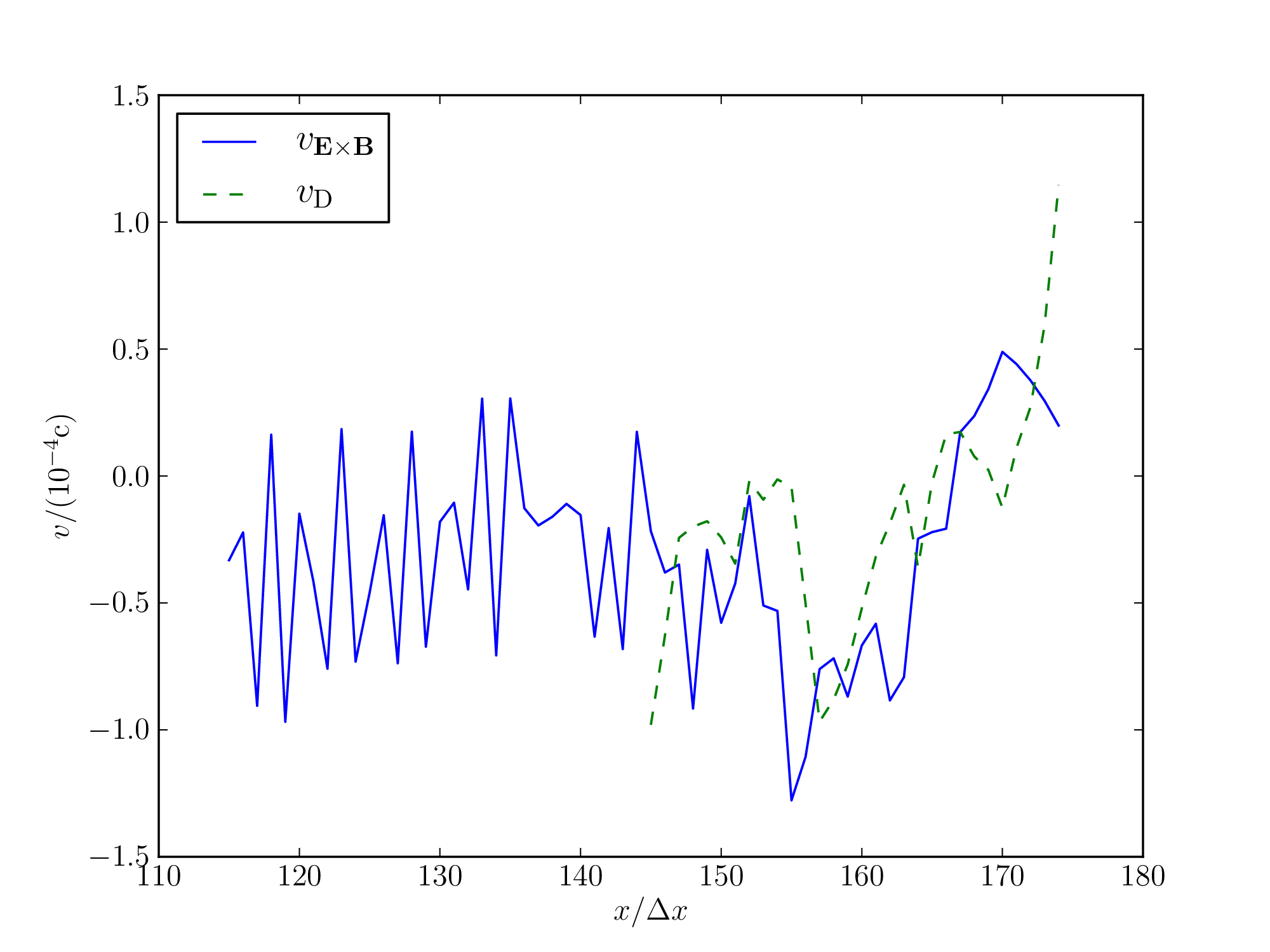} 
\par\end{centering}
}
\par\end{centering}
\caption{The toroidal averaged electric field in $\bfe_{x}$ direction (a).
The $E\times B$ velocity $v_{E\times B}$ and measured phase velocity
$v_{D}$ for the $n=12$ mode in the $\bfe_{z}$ direction at $z=128\Delta x$
and $t=8a/v_{t,i}$ (b) .}
\label{FigExB} 
\end{figure}

The time-history of the mode amplitude is shown in Fig.~\ref{FigModeAMP}.
The unstable mode saturates approximately at $t\approx5a/v_{t,i}$,
and the saturation level is in the range of 2\%. Recent nonlinear
gyrokinetic simulation \citep{dong2019nonlinear} suggested that the
instability is saturated by the $E\times B$ zonal flow generated
by the instability. To verify this mechanism in our 6D fully kinetic
simulation, the toroidally averaged $\bfE$, $E\times B$ velocity
and the measured phase velocity of the $n=12$ mode at $z=128\Delta x$
and $t=8a/v_{t,i}$ are compared in \FIG{FigExB}. The $E\times B$
velocity at the edge correlates strongly with the phase velocity of
the perturbation in terms of amplitude and profile. As a result, the
$E\times B$ flow for the background plasma generated by instability
interacts coherently with the mode structure, significantly modifies
the space-time structure of the perturbation relative to the background
plasma and reduces the drive of the instability. For this case simulated,
the nonlinear saturation mechanism agrees qualitatively with the nonlinear
gyrokinetic simulation \citep{dong2019nonlinear}.

\section{Conclusions}

\label{Sec4} Even though 6D kinetic PIC method is a classical simulation
tool for plasma physics, up to now it has not been applied to numerical
studies of tokamak physics in spite of continuous improvement \citep{okuda1972,cohen1982a,langdon1983,cohen1989a,liewer1989,friedman1991,eastwood1991virtual,cary1993,villasenor1992rigorous,qin00-084401,qin00-389,qin01-477,Davidson01-all,esirkepov2001exact,vay2002,nieter2004vorpal,huang2006,Crouseilles07,chen2011,Chacon2013,evstatiev2013,Shadwick14,Moon2015,Huang2016,xiao2019commet,Webb2016,Li2019,Li2020,Holderied2020,Zheng2020,wang2020geometric,Kormann2021}.
In the present study, we have developed an explicit structure-preserving
geometric PIC algorithm in curvilinear orthogonal meshes, in particular
the cylindrical mesh, and apply it to carry out whole-device 6D kinetic
simulation studies of tokamak physics. The work reported represents
a further development of the structure-preserving geometric PIC algorithm
\citep{squire4748,squire2012geometric,xiao2013variational,xiao2015explicit,xiao2015variational,he2015hamiltonian,qin2016canonical,he2016hamiltonian,kraus2017gempic,xiao2017local,xiao2018structure,xiao2019field},
achieving the goal of practical applications in magnetic fusion research.

Along with it predecessors \citep{squire4748,squire2012geometric,xiao2013variational,xiao2015explicit,xiao2015variational,he2015hamiltonian,qin2016canonical,he2016hamiltonian,kraus2017gempic,xiao2017local,xiao2018structure,xiao2019field},
the algorithm extends the symplectic integration method for finite
dimensional canonical Hamiltonian systems developed since the 1980s
\citep{Devogelaere56,lee82,Ruth83,Feng85,Feng86,lee87,SanzSerna1988,veselov88,yoshida1990construction,Forest90,Channell90,Candy91,Tang93,Sanz-Serna94,Shang94,Feng95,Shang99,marsden2001discrete,Guo2002,Hairer02, Hong02,Shang2006,Feng10,zhang2016explicit,Tao2016},
and preserves an infinite dimensional non-canonical symplectic structure
of the particle-field systems. In addition, other important geometric
structures and conservation laws, such as the gauge symmetry, the
local charge conservation law \citep{squire4748,squire2012geometric,xiao2018structure,glasser2019b}
and the local energy conservation law \citep{xiao2017local}, are
preserved exactly as well. These preserved structures and conservation
laws improve the accuracy and fidelity of large-scale long-term simulations
on modern computing hardware \citep{fu2016sunway}.

Through the whole-device 6D kinetic simulation, we numerically obtained
a self-consistent kinetic steady state for fusion plasma in the tokamak
geometry. It was found that the pressure tensor of the self-consistent
kinetic steady state is diagonal, anisotropic in 3D, but isotropic
in the poloidal plane. The steady state also includes a steady-state
sub-sonic ion flow in the range of $10$km/s, which agrees with previous
experimental observations \citep{Ince-Cushman2009,Rice2009} and theoretical
calculations \citep{Guan2013,Guan2013a}. Kinetic ballooning instability
in the self-consistent kinetic steady state was successfully simulated.
In the linear phase, it was found that high-$n$ ballooning modes
have larger growth rates than low-$n$ global modes. In the nonlinear
phase, the modes saturate approximately in $5$ ion transit times
at the $2$\% level by the $E\times B$ flow generated by the instability.
These results qualitatively agrees with early \citep{Qin98-thesis,Qin1999}
and recent \citep{dong2019nonlinear} simulations by electromagnetic
gyrokinetic codes. In addition, compared with conventional gyrokinetic
and reduced Braginskii \citet{zeiler1997nonlinear,PhysRevLett.105.175005,ricci2012simulation}
fluid simulation methods, more physical effects, such as fully kinetic
dynamics and the self-consistent radial electric field, are naturally
included in the present method. These effects can be crucial for edge
plasmas and will be investigated in the next step.

It worth mentioning that in the present work we can not directly control
the 2D kinetic steady state because it is numerically evolved from
a given initial condition. In the future, we plan to solve this problem
by adopting MHD equilibrium solutions as the initial conditions. Because
a MHD equilibrium should be at least close to a lowest order kinetic
steady state, it is expected that a kinetic steady state can be obtained
by a short time evolution.

The present work can be also extended to describe more complex physical
processes in tokamak plasmas. For example, we can add energetic particles
to investigate their interactions with the background plasma. An antenna
can be also modeled as a current source to study the wave heating
and current drive \citep{zheng2020structure}. To simulate collision
related physics, we can include Monte-Carlo Collision (MCC) \citep{birdsall1991particle}
processes. It should be noted that due to the lack of marker particles
in PIC simulations, the numerical collision frequency is usually larger
than the real collision frequency of the plasma. More investigations
are needed to determine the proper method to simulate collisions in
the present scheme. Adding more physical effects to the geometric
structure preserving PIC simulation framework will help us to better
understand the tokamak physics. These topics will be addressed in
the future study.

\appendix

\section{External magnetic field of the tokamak, initial particle loading,
and the boundary setup}

\label{SecAPP} In this appendix we describe the setup of external
magnetic field of the tokamak, initial particle loading, and the boundary
setup for the simulation study. The normalization of quantities are
listed in Table \ref{TabUNIT}. 
\begin{table}
\centering %
\begin{tabular}{|c|c|c|}
\hline 
Physical quantity  & Symbol(s)  & Unit\tabularnewline
\hline 
Length  & $x,r,\dots$  & $\Delta x$\tabularnewline
\hline 
Velocity  & $\bfv$  & $\rmc_{n}$\tabularnewline
\hline 
Mass  & $m_{i}$  & $m_{e}$\tabularnewline
\hline 
Time  & $t$  & $\Delta x/\rmc_{n}$\tabularnewline
\hline 
\end{tabular}\caption{Normalization for quantities.}
\label{TabUNIT} 
\end{table}

The magnetic field is divided into three parts, 
\begin{eqnarray}
\bfB_{0,\textrm{init}}=\bfB_{0,p}+\bfB_{0,e}+\bfB_{0,t},
\end{eqnarray}
where $\bfB_{0,e}$ is the external magnetic fields generated by poloidal
coils, $\bfB_{0,t}$ is the magnetic field generated by toloidal coils,
i.e., 
\begin{equation}
\bfB_{0,t}=\frac{B_{0}R_{0}}{R}\thinspace,
\end{equation}
and they do not evolve with time. $\bfB_{0,p}$ is the magnetic field
generated by the plasma current. The current $\bfJ$ and the vector
potential $\bfA$ are related by 
\begin{eqnarray}
\nabla\times\left(\nabla\times\bfA\right) & = & \bfJ~.\label{EqnMAXJ}
\end{eqnarray}
Initially, the current is in the $y$-direction and depends only on
$x$ and $z$. In the adopted cylindrical coordinate $(x,y,z)$ the
line element is 
\begin{eqnarray}
\rmd s^{2} & = & \left(\rmd x\right)^{2}+\left(\frac{x+R_{0}}{R_{0}}\rmd y\right)^{2}+\left(\rmd z\right)^{2}~,
\end{eqnarray}
and Eq.\,(\ref{EqnMAXJ}) becomes 
\begin{eqnarray}
\frac{\partial^{2}}{\partial z^{2}}A_{y}\left(x,z\right)+\frac{\partial}{\partial x}\left(\frac{\partial}{\left(x+R_{0}\right)\partial x}\left(A_{y}\left(x,z\right)\left(x+R_{0}\right)\right)\right) & = & -j_{y}\left(x,z\right)~,\\
\left(\frac{\partial^{2}}{\partial x^{2}}+\frac{\partial^{2}}{\partial z^{2}}\right)A_{y}^{o}\left(x,z\right)-\frac{1}{x+R_{0}}\frac{\partial}{\partial x}A_{y}^{o}\left(x,z\right) & = & -j_{y}^{o}\left(x,z\right)~,\label{EqnDIFCUR}
\end{eqnarray}
where 
\[
f^{o}(x,z)=f(x,z)\frac{x+R_{0}}{R_{0}}~.
\]
When $j_{y}\left(x,z\right)=I_{0}\delta\left(x-x_{0}\right)\delta\left(z-z_{0}\right)$,
which represents a coil current at $(x_{0},z_{0})$, \EQ{EqnDIFCUR}
can be solved using spherical harmonic expansion. However the convergence
of the series is slow when $r_{g}=\sqrt{\left(x+r_{0}\right)^{2}+(z-z_{0})^{2}}$
approaches $x_{0}+R_{0}$, the radius of the coil. We note that the
second term in the left-hand-side of \EQ{EqnDIFCUR} is negligible
when $R_{0}+x\gg B_{z}/\left(\partial B_{z}/\partial x\right)$. In
this case, \EQ{EqnDIFCUR} simplifies to 
\begin{eqnarray}
\left(\frac{\partial^{2}}{\partial x^{2}}+\frac{\partial^{2}}{\partial z^{2}}\right)A_{y}^{o}\left(x,z\right) & = & -j_{y}^{o}\left(x,z\right)~,\label{EqnDIFCUR1}
\end{eqnarray}
which is a standard 2D Poisson equation. Its solution for the coil
current at $(x_{0},z_{0})$ with $I_{0}=1$ is 
\begin{eqnarray}
A_{y,PF}^{o}(x,z;x_{0},z_{0}) & = & \frac{1}{4\pi}\log\left(\left(x-x_{0}\right)^{2}+\left(z-z_{0}\right)^{2}\right)~.\label{EqnAY0}
\end{eqnarray}
Here, dimensionless variables have been used to simplify the notation.
The total external vector potential generated by poloidal field coils
is 
\begin{eqnarray}
A_{y,0,e}^{o}\left(x,z\right)=\sum_{i}J_{PF,i}A_{y,PF}^{o}\left(x,z;x_{PF,i},z_{PF,i}\right)~,
\end{eqnarray}
where the locations of poloidal field coils are displayed in Tab.~\ref{TabLPFS}.
\begin{table}
\centering %
\begin{tabular}{|c|c|c|c|c|c|}
\hline 
Coil number $i$  & $x/$m  & $z/$m  & Coil number $i$  & $x/$m  & $z/$m\tabularnewline
\hline 
1  & -0.05  & 0.4896  & 7  & 0.48  & -0.1904\tabularnewline
\hline 
2  & -0.05  & 0.3296  & 8  & 0.48  & 1.0096\tabularnewline
\hline 
3  & -0.05  & 0.6496  & 9  & 0.62  & 0.0156\tabularnewline
\hline 
4  & -0.05  & 0.1696  & 10  & 0.62  & 0.8036\tabularnewline
\hline 
5  & -0.05  & 0.8096  & 11  & 0.08  & 1.0096\tabularnewline
\hline 
6  & -0.05  & 0.0096  & 12  & 0.08  & -0.1904\tabularnewline
\hline 
\end{tabular}\caption{Locations of tokamak poloidal field coils.}
\label{TabLPFS} 
\end{table}

For $\bfB_{0,p}$ and the corresponding plasma current $\bfJ_{0}$,
we first construct a vector potential $\bfA_{y,0,p}$ and then use
this potential to obtain $\bfB_{0,p}$ and $\bfJ_{0}$. The constructed
$A_{y,0,p}$ is 
\[
A_{y,0,p}^{o}\left(x,z\right)=\left\{ \begin{array}{cc}
-\frac{r^{2}B_{0}}{2q_{0}r_{0}}~, & \textrm{ }r\leq r_{l}\,,\\
-\frac{B_{0}}{q_{0}r_{0}}\frac{9r^{2}r_{r}+6r_{l}^{3}\log(r_{l})+\left(-6\log(r)-5\right)r_{l}^{3}-4r^{3}}{18r_{r}-18r_{l}}~, & r_{l}<r\leq r_{r}\,,\\
-\frac{B_{0}}{q_{0}r_{0}}\frac{6r_{r}^{3}\log(r_{r})+\log(r)\left(6r_{r}^{3}-6r_{l}^{3}\right)+5r_{r}^{3}+6r_{l}^{3}\log(r_{l})-5r_{l}^{3}}{18r_{r}-18r_{l}}~, & \textrm{otherwise},
\end{array}\right.~
\]
where $r=\sqrt{\left(x-x_{\mathrm{mid}}\right)^{2}+\left(z-z_{\mathrm{mid}}\right)^{2}}$,
$x_{\mathrm{mid}}=N_{x}/2$ and $z_{\mathrm{mid}}=N_{z}/2$ are coordinates
of the center of simulation domain, $q_{0}$ is the safety factor
in the core of the plasma, $r_{l}=0.454a=0.1$m and $r_{r}=0.667a=0.147$m
are two parameters that determine the current density distribution.
The discrete magnetic fields are obtained by 
\begin{eqnarray}
\bfB_{J,0,p/e} & = & \CURLD\bfA_{0,p/e}\left(i,j,k\right)~,
\end{eqnarray}
and the discrete current density is obtained from 
\begin{eqnarray}
\bfJ_{J,0,\mathrm{all}} & = & \CURLDP\left(\left(\bfB_{J,0,p}+\bfB_{J,0,e}\right)\left[\left(1+i/r_{0}\right)^{-1},1+i/r_{0},\left(1+i/r_{0}\right)^{-1}\right]\right)~.
\end{eqnarray}
The final plasma current is chosen as 
\begin{eqnarray}
\bfJ_{J,0}=\left\{ \begin{array}{cc}
\bfJ_{J,0,\mathrm{all}}~, & \textrm{when }r<r_{r}~,\\
0~, & \textrm{otherwise.}
\end{array}\right.~
\end{eqnarray}

Density and temperature are calculated from $A_{y,0}^{o}=A_{y,0,e}^{o}+A_{y,0,p}^{o}$.
We introduce a reference function $g_{r}$ defined as 
\begin{equation}
g_{r}(x,z)=1-\left\{ \begin{array}{cc}
0~, & A_{y,0}^{o}\left(x,z\right)<A_{y,\min}^{o}\,,\\
\frac{A_{y,0}^{o}\left(x,z\right)-A_{y,\min}^{o}}{A_{y,\max}^{o}-A_{y,\min}^{o}}~, & A_{y,\min}^{o}\leq A_{y,0}^{o}\left(x,z\right)<A_{y,\max}^{o}\,,\\
1~, & A_{y,\max}^{o}\leq A_{y,0}^{o}\,,
\end{array}\right.
\end{equation}
where 
\begin{eqnarray}
A_{y,\min}^{o} & = & A_{y,0}^{o}\left(0.267N_{x},0.114N_{z}\right)~,\\
A_{y,\max}^{o} & = & A_{y,0}^{o}\left(0.488N_{x},0.369N_{z}\right)~.
\end{eqnarray}
The initial density and temperature for electrons and ions are 
\begin{eqnarray}
n_{e}=n_{i} & = & n_{e,0}\left(\left(g_{r}\left(x,z\right)^{2}+1\right)\left(1-g_{r}\left(x,z\right)^{2}\right)\right)^{3}~,\\
T_{e}=T_{i} & = & T_{e,0}\left(\left(g_{r}\left(x,z\right)^{2}+1\right)\left(1-g_{r}\left(x,z\right)^{2}\right)\right)^{6}~.
\end{eqnarray}
In this work, ions are all deuterium ions, and the initial velocity
distribution for each specie is Maxwellian with a flow velocity, i.e.,
\begin{eqnarray}
f_{e/i,0}\left(\bfx,\bfv\right) & = & \frac{n_{e/i}\left(x,z\right)}{\left(2\pi T_{e/i}/m_{e/i}\right)^{3/2}}\exp\left(-\frac{|\bfv+v_{e/i,y,0}\left(x,z\right)\bfe_{y}|^{2}}{2T_{e/i}/m_{e/i}}\right)~,
\end{eqnarray}
where 
\begin{eqnarray}
v_{e/i,y,0}\left(x,z\right)=\frac{j_{y,0}\left(x,z\right)m_{i/e}}{q_{e/i}\left(m_{i}-m_{e}\right)n_{e/i}\left(x,z\right)}~.
\end{eqnarray}

The boundaries of the simulations are configured as follows. The boundaries
at $x=0$, $x=N_{x}\Delta x$, $z=0$, and $z=N_{z}\Delta z$ are
chosen to be perfect electric conductors for the electromagnetic fields.
For particles, we introduce a thin slow-down layer, 
\begin{align*}
\mathbf{L}_{\mathrm{p}}= & \big\{\left(x,y,z\right)|0<x<5\Delta x\quad\textrm{or}\quad0<z<5\Delta z\quad\textrm{or}\\
 & \left(N_{x}-5\right)\Delta x<x<N_{x}\Delta x\quad\textrm{or}\quad\left(N_{z}-5\right)\Delta z<z<N_{z}\Delta z\Big\}~.
\end{align*}
If a particle is inside $\mathbf{\mathbf{L}_{\mathrm{p}}}$ at a time-step,
its velocity $\bfv_{p}$ will be reduced to $0.98\bfv_{p}$ at the
end of this time-step, and the particle will be removed when $\bfv_{p}$
is smaller than $1\EXP{-5}\rmc$. Periodic boundaries for both particles
and electromagnetic fields are adopted in the $y$-direction. The
plasma is confined inside the last closed flux surface mostly, and
the shape of plasma is not directly related to the simulation boundaries.

\section{Explicit 2nd-order structure-preserving geometric PIC algorithm in
the cylindrical mesh}

In this appendix, we list the detailed update rule for the 2nd-order
structure-preserving geometric PIC algorithm in the cylindrical mesh
introduced in Sec. \ref{Sec2cylindrical}. It updates previous particle
locations and discrete electromagnetic fields $\{\bfx_{sp,2l-2},\bfx_{sp,2l-1},\bfE_{J,l},\bfB_{K,l}\}$
to the current ones $\{\bfx_{sp,2l},\bfx_{sp,2l+1},\bfE_{J,l+1},\bfB_{K,l+1}\}$,

\begin{eqnarray*}
x_{sp,2l} & = & 2x_{sp,2l-1}-x_{sp,2l-2}+\Delta x^{-2}\left(2\frac{q_{s}}{m_{s}}h^{2}E_{x,1}\right)~,\\
y_{sp,2l} & = & \left(r_{0}^{2}+2r_{0}x_{sp,2l}+x_{sp,2l}^{2}\right)^{-1}\left(\left(-2r_{0}x_{sp,2l-2}-r_{0}^{2}-x_{sp,2l-2}^{2}\right)y_{sp,2l-2}+\right.\\
 &  & \left.\left(2r_{0}^{2}+2r_{0}x_{sp,2l-2}+x_{sp,2l-2}^{2}+2r_{0}x_{sp,2l}+x_{sp,2l}^{2}\right)y_{sp,2l-1}\right)+\\
 &  & r_{0}^{2}\left(\frac{q_{s}}{m_{s}}h\left(2E_{y,1}h-\left(B_{z,0,0,1,1,0}+B_{z,1,1,1,1,0}\right)\right)\right)\left(\Delta y^{2}r_{0}^{2}+2\Delta y^{2}r_{0}x_{sp,2l}+\Delta y^{2}x_{sp,2l}^{2}\right)^{-1}~,\\
z_{sp,2l} & = & 2z_{sp,2l-1}-z_{sp,2l-2}+\\
 &  & \Delta z^{-2}\left(\frac{q_{s}}{m_{s}}h\left(2E_{z,1}h+\left(B_{y,0,0,1,1,0}-B_{x,0,0,0,1,1}-B_{x,1,2,1,1,1}+B_{y,1,1,1,1,0}\right)\right)\right)~,\\
z_{sp,2l+1} & = & 2z_{sp,2l}-z_{sp,2l-1}~,\\
y_{sp,2l+1} & = & 2y_{sp,2l}-y_{sp,2l-1}+r_{0}^{2}\left(h\frac{q_{s}}{m_{s}}\left(B_{x,1,2,2,2,2}+B_{x,1,2,2,1,2}\right)\right)\left(\Delta y^{2}r_{0}^{2}+2\Delta y^{2}r_{0}x_{sp,2l}+\Delta y^{2}x_{sp,2l}^{2}\right)^{-1}\\
x_{sp,2l+1} & = & \Delta x^{-2}r_{0}^{-2}\left(-\Delta x^{2}r_{0}^{2}x_{sp,2l-1}+2\Delta x^{2}r_{0}^{2}x_{sp,2l}+\left(\Delta y^{2}r_{0}+\Delta y^{2}x_{sp,2l}\right)y_{sp,2l-1}^{2}\right.\\
 &  & \left.+\left(-2\Delta y^{2}r_{0}-2\Delta y^{2}x_{sp,2l}\right)y_{sp,2l-1}y_{sp,2l}+\left(2\Delta y^{2}r_{0}+2\Delta y^{2}x_{sp,2l}\right)y_{sp,2l}^{2}\right.\\
 &  & \left.+\left(-2\Delta y^{2}r_{0}-2\Delta y^{2}x_{sp,2l}\right)y_{sp,2l}y_{sp,2l+1}+\left(\Delta y^{2}r_{0}+\Delta y^{2}x_{sp,2l}\right)y_{sp,2l+1}^{2}\right)~,\\
 &  & +\Delta x^{-2}\left(-h\frac{q_{s}}{m_{s}}\left(B_{y,1,2,2,1,2}+B_{y,1,2,2,2,2}-B_{z,1,2,1,1,1}-B_{z,1,2,2,3,1}\right)\right)~,\\
\end{eqnarray*}
\begin{eqnarray*}
\frac{1}{\bfha\left(\bfx_{J}\right)\bfha\left(\bfx_{J}\right)}\frac{\bfE_{J,l+1}-\bfE_{J,l}}{\Delta t}\bfhc\left(\bfx_{J}\right) & = & \sum_{K}\CURLDP_{J,K}\frac{\bfhc\left(\bfx_{J}\right)}{\bfhb\left(\bfx_{K}\right)\bfhb\left(\bfx_{K}\right)}\bfB_{K,l}-\bfJ_{J,l}~,\\
\frac{\bfB_{K,l+1}-\bfB_{K,l}}{\Delta t} & = & -\sum_{J}\CURLD_{K,J}\bfE_{J,l+1}~,
\end{eqnarray*}
where 
\begin{eqnarray*}
h & = & \Delta t/2~,\\
\bfB_{t,i_{1},i_{2},i_{3},i_{4}} & = & \left[\begin{array}{c}
B_{x,t,i_{1},i_{2},i_{3},i_{4}}\\
B_{y,t,i_{1},i_{2},i_{3},i_{4}}\\
B_{z,t,i_{1},i_{2},i_{3},i_{4}}
\end{array}\right]=\int_{x_{i_{4}+1,sp,2l+i_{i_{4}+1}}}^{x_{i_{4}+1,sp,2l+i_{i_{4}+1}+1}}\rmd\bar{x}\sum_{J}\bfB_{J,l+t-1}\\
 &  & \WTWO{\left[x_{sp,2l+i_{1}},y_{sp,2l+i_{2}},z_{sp,2l+i_{3}}\right]+\bfe_{x_{i_{4}+1}}\left(\bar{x}-x_{i_{4}+1,sp,2l+i_{i_{4}+1}}\right)}~,\\
\bfE_{1} & = & \left[\begin{array}{c}
E_{x,1}\\
E_{y,1}\\
E_{z,1}
\end{array}\right]=\sum_{J}\bfE_{J,l}\WONE{\bfx_{sp,2l-1}}~,\\
\bfJ_{J,l} & = & \frac{1}{\Delta t}\sum_{s,p}q_{s}\int_{C_{sp,2l-1,2l}\bigcup C_{sp,2l,2l+1}^{*}}\WONE{\bfx}\rmd\bfx~,\\
C_{sp,2l-1,2l} & = & \left\{ \left(x_{1,sp2l-1}+\tau\left(x_{1,sp,2l}-x_{1,sp2l-1}\right),x_{2,sp2l-1},x_{3,sp2l-1}\right)|\tau\in\left[0,1\right)\right\} \bigcup\\
 &  & \left\{ \left(x_{1,sp,2l},x_{2,sp2l-1}+\tau\left(x_{2,sp,2l}-x_{2,sp2l-1}\right),x_{3,sp2l-1}\right)|\tau\in\left[0,1\right)\right\} \bigcup\\
 &  & \left\{ \left(x_{1,sp,2l},x_{2,sp,2l},x_{3,sp2l-1}+\tau\left(x_{3,sp,2l}-x_{3,sp2l-1}\right)\right)|\tau\in\left[0,1\right)\right\} ~,\\
C_{sp,2l,2l+1}^{*} & = & \left\{ \left(x_{1,sp,2l}+\tau\left(x_{1,sp,2l+1}-x_{1,sp,2l}\right),x_{2,sp,2l+1},x_{3,sp,2l+1}\right)|\tau\in\left[0,1\right)\right\} \bigcup\\
 &  & \left\{ \left(x_{1,sp,2l},x_{2,sp,2l}+\tau\left(x_{2,sp,2l+1}-x_{2,sp,2l}\right),x_{3,sp,2l+1}\right)|\tau\in\left[0,1\right)\right\} \bigcup\\
 &  & \left\{ \left(x_{1,sp,2l},x_{2,sp,2l},x_{3,sp,2l}+\tau\left(x_{3,sp,2l+1}-x_{3,sp,2l}\right)\right)|\tau\in\left[0,1\right)\right\} ~.
\end{eqnarray*}

\label{SecESPIC}

\section*{Acknowledgment}

J. Xiao was supported by the the National MCF Energy R\&D Program
(2018YFE0304100), National Key Research and Development Program (2016YFA0400600,
2016YFA0400601 and 2016YFA0400602), and the National Natural Science
Foundation of China (NSFC-11905220 and 11805273). J. Xiao developed
the algorithm and the \textsl{SymPIC} code and carried out the simulation
on Tianhe 3 prototype at the National Supercomputer Center in Tianjin
and Sunway Taihulight in the National Supercomputer Center in Wuxi.
H. Qin was supported by the U.S. Department of Energy (DE-AC02-09CH11466).
H. Qin contributed to the physical study of the self-consistent kinetic
equilibrium and the kinetic ballooning modes. 

 \bibliographystyle{apsrev4-1}
\bibliography{xp}

\end{document}